\documentclass[apl,10pt,twocolumn,noshowpacs,superscriptaddress]{revtex4-2}

\usepackage{graphicx}
\usepackage{multirow}
\usepackage{color}
\usepackage{amsmath}
\usepackage{bm}
\usepackage{xcolor}
\usepackage{braket}
\usepackage{siunitx}
\usepackage{hyperref}
\usepackage{hyperxmp}

\usepackage{subfigure}

\renewcommand{\vec}[1]{\boldsymbol{#1}}

\newcommand{\bk}{\vec{k}}

\newcommand{\energy}{\mathcal{E}}
\newcommand{\fermi}{\energy_{\text{F}}}

\usepackage[
    type={CC},
    modifier={by},
    version={3.0},
]{doclicense}

\begin{document}

\title{Current-induced spin and orbital polarization in the ferroelectric Rashba semiconductor GeTe}

\author{Sergio~Leiva-Montecinos}
\email{sergio-tomas.leiva-montecinos@physik.uni-halle.de}
\affiliation{Institut für Physik, Martin-Luther-Universität Halle-Wittenberg, D-06099 Halle (Saale), Germany}

\author{Libor~Vojáček}
\affiliation{Univ. Grenoble Alpes, CEA, CNRS, SPINTEC, 38054 Grenoble, France}

\author{Jing~Li}
\affiliation{Université Grenoble Alpes, CEA, Leti, F-38000, Grenoble, France}

\author{Mairbek~Chshiev}
\affiliation{Univ. Grenoble Alpes, CEA, CNRS, SPINTEC, 38054 Grenoble, France}
\affiliation{Institut Universitaire de France, Paris, 75231, France}

\author{Laurent~Vila}
\affiliation{Univ. Grenoble Alpes, CEA, CNRS, SPINTEC, 38054 Grenoble, France}

\author{Ingrid~Mertig}
\affiliation{Institut für Physik, Martin-Luther-Universität Halle-Wittenberg, D-06099 Halle (Saale), Germany}

\author{Annika~Johansson}
\affiliation{Max Planck Institute of Microstructure Physics, Weinberg 2, 06120 Halle (Saale), Germany}

\begin{abstract}

The Edelstein effect is a promising mechanism for generating spin and orbital polarization from charge currents in systems without inversion symmetry. In ferroelectric materials, such as Germanium Telluride (GeTe), the combination of bulk Rashba splitting and voltage-controlled ferroelectric polarization provides a pathway for electrical control of the sign of the charge-spin conversion. 
In this work, we investigate current-induced spin and orbital magnetization in bulk GeTe using Wannier-based tight-binding models derived from \textit{ab initio} calculations and semiclassical Boltzmann theory. Employing the modern theory of orbital magnetization, we demonstrate that the orbital Edelstein effect entirely dominates its spin counterpart. This difference is visualized through the spin and orbital textures at the Fermi surfaces, where the orbital moment surpasses the spin moment by one order of magnitude. Moreover, the orbital Edelstein effect remains largely unaffected in the absence of spin-orbit coupling, highlighting its distinct physical origin compared to the spin Edelstein effect.
\end{abstract}

\maketitle

\section{Introduction}

The pursuit of electric control of magnetic properties has led to intense exploration of materials exhibiting Rashba-like spin splitting in their electronic structures. Rashba spin-orbit coupling, originally discussed for three-dimensional wurtzite crystals~\cite{rashba2015symmetry, Bihlmayer2015} and later for two-dimensional systems with broken inversion symmetry~\cite{Rashba1960, Rashba1984, Rashba1984_2}, can also occur in the bulk of ferroelectric semiconductors, called ferroelectric Rashba semiconductors (FERSC)~\cite{DiSante_2013, Picozzi2014, Sharma2021, Gu2024}. Among them, Germanium Telluride (GeTe) is a paradigm material example. The switchable ferroelectric polarization allows electric control of the Rashba-like band splitting, and hence also the charge-spin conversion~\cite{DiSante_2013, Rinaldi_2016, Rinaldi_2018, Varotto_2021, Vojek2024}.

\begin{figure*}
\centering
\includegraphics[width=1\linewidth]{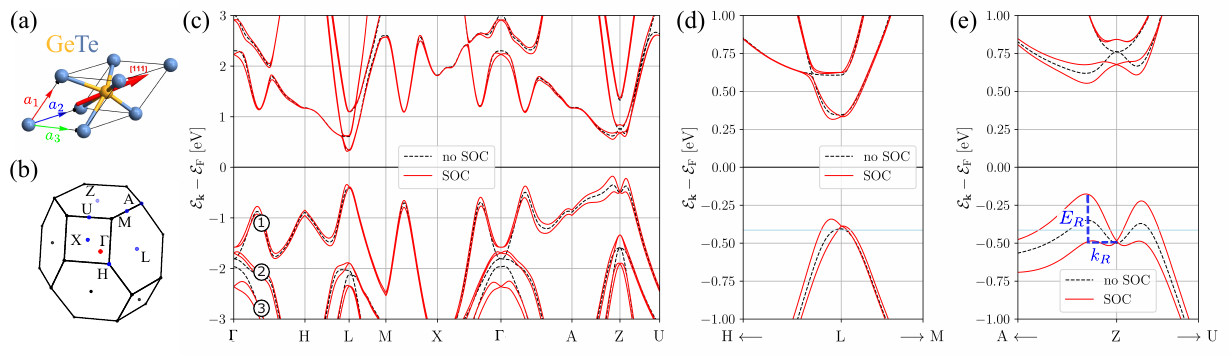}
\caption{Electronic structure of $\alpha$-GeTe. (a) Crystal structure of GeTe with the primitive lattice vectors, and a relative displacement of the central Ge atom along the [111] direction, indicated by the red arrow. (b) First Brillouin zone with labeled high-symmetry points of the distorted rhombohedral structure. (c) Band structure from the Wannier tight-binding model with (solid lines) and without (dashed lines) spin-orbit coupling (SOC), along the high-symmetry path. (d-e) Zoom-in energy around the (d) $L$ and (e) $Z$ point, including the effective Rashba energy splitting $E_R$ and momentum splitting $k_R$. In panel (c), the valence bands are grouped in three pairs \textcircled{\raisebox{-0.9pt}{$1$}}, \textcircled{\raisebox{-0.9pt}{$2$}}, \textcircled{\raisebox{-0.9pt}{$3$}} for further analysis.}
\label{fig:electronic_structure}
\end{figure*}

One of the most promising phenomena providing charge-spin conversion is the spin Edelstein effect (SEE) \cite{aronov1989nuclear, edelstein1990spin}, also known as the inverse spin galvanic effect, or current-induced spin polarization. An applied external electric field generates a homogeneous spin polarization in systems without inversion symmetry due to spin-orbit coupling (SOC). This effect has been studied in many materials with various sources of broken inversion symmetry, such as Rashba systems \cite{LeivaMontecinos2023, fontenele2024impact, Johansson2016, Gaiardoni2025, dey2025current}, topological insulators~\cite{Culcer2010, Mellnik2010, Ando2014, Kondou2016, Johansson2016, Rojas2016}, Weyl semimetals \cite{johansson2018edelstein, Zhao2020}, chiral materials \cite{Shalygin2012, Furukawa2017, Furukawa2021, yang2021chiral, Calavalle2022, gobel2025chirality}, oxide interfaces \cite{Vaz2019, Johansson2021, trama2022tunable, varotto2022direct, krishnia2024interfacial}, twisted heterostructures \cite{katsantonis2023giant, orazbay2024twistronics}, p-wave magnets \cite{pari2025nonrelativistic, chakraborty2024highly}, transition metal dichalcogenides \cite{Cysne2021nano, Lee2022, InglaAyns2022, Cysne2023}, ferroelectric semiconductors \cite{Varotto_2021, Chirolli2022}, and other quantum materials \cite{Han2018}.

A second contribution to the Edelstein effect, the orbital Edelstein effect (OEE), arises from the orbital angular momentum of the electrons~\cite{thonhauser2011theory, Park2011, Park2012, Go2021, Johansson2024, BurgosAtencia2024, cysne2025orbitronics}, resulting in a current-induced nonequilibrium orbital magnetization in systems with broken inversion symmetry. Unlike the SEE, the OEE does not require the presence of SOC, as has been shown for various magnetic and nonmagnetic materials and their interfaces~\cite{yoda2015current, yoda2018orbital, Go2017, Salemi2019, Nikolaev2024}.

Calculating the orbital angular momentum in extended bulk crystals comes with the challenge of defining the position operator of translationally invariant systems. To circumvent this problem, the modern theory of orbital magnetization (MTOM) has been introduced, using localized Wannier functions and wave packets from Bloch states, respectively. In contrast to the atom-centered approximation (ACA), within which the orbital angular momentum is evaluated in finite spheres around the atomic positions, the MTOM also includes non-local, itinerant contributions to orbital magnetism~\cite{thonhauser2011theory}. 
The MTOM has been implemented in several density functional theory (DFT) codes~\cite{lopez2012wannier, mostofi_updated_2014, Tsirkin2021} and tight-binding models~\cite{yoda2015current, yoda2018orbital}. 
Previous experimental studies on GeTe have mostly focused on spin textures of the surface states and bulk states~\cite{Krempasky2016, Krempasky2016_2, Krempasky2020} of GeTe, and on charge-spin conversion via the inverse spin Hall effect and the inverse SEE~\cite{Rinaldi_2016, Varotto_2021}. Spin textures~\cite{DiSante_2013} of GeTe and the resulting SEE~\cite{Tenzin2023} and spin Hall effect~\cite{Varotto_2021} have been investigated theoretically. In Ref.~\cite{Ponet2018}, orbital textures of GeTe are calculated in selected regions of the Brillouin zone, using the ACA, and a giant orbital Rashba splitting is identified.  However, in addition to these existing studies, we analyze the orbital textures and the OEE in GeTe using the MTOM and demonstrate that the orbital contribution significantly dominates the total Edelstein effect.

In this work, we calculate both charge-spin and charge-orbital conversion via the SEE and OEE, respectively, in bulk GeTe by employing a multiscale approach comprising DFT, a Wannier tight-binding model, the MTOM, and semiclassical transport theory. The paper is organized as follows: In Section~\ref{sec:methods}, we provide an overview of the computational methods and the theory of the SEE and OEE. In Section~\ref{sec:elec_str}, we discuss the electronic structure with an estimation of the Rashba parameter. In Section~\ref{sec:spin_orbital}, we present the spin and orbital textures and Edelstein effects in GeTe.%

\section{Methodology}\label{sec:methods}

Density functional theory (DFT) calculations were performed using the Vienna \textit{Ab initio} Simulation Package (\texttt{VASP})~\cite{Kresse_1993, Kresse_1996a, Kresse_1996b, Kresse_1999} with Perdew–Burke-Ernzerhof (PBE) functional \cite{Perdew_1996} and DFT-D3 correction of Grimme~\textit{et al.}~\cite{Grimme_2010} with Becke-Johnson damping~\cite{Grimme2011}.
The Brillouin zone of bulk GeTe is sampled by a $12\times 12 \times 12 $ $\Gamma$-centered $k$-mesh. A plane-wave basis set is used with a cut-off energy of $\SI{520}{\electronvolt}$.
The convergence for the electronic density is set to $10^{-7}\si{\electronvolt}$. The atomic structure is relaxed with the threshold forces of $\SI{1}{\milli\electronvolt\per\angstrom}$. 

The Wannier tight-binding models were constructed using the \texttt{Wannier90}~package~\cite{marzari_maximally_1997,marzari_maximally_2012, mostofi_updated_2014} including the spin operator~\cite{vojacek_field-free_2024,vojacek_multiscale_2024} for a 16-orbital basis (Ge,Te-s,p orbitals) within a $9 \times 9 \times 9$ real-space supercell. The orbital angular momentum was interpolated in $k$-space using the \texttt{wannierBerri} package~\cite{Tsirkin2021}.

The magnetic moment per unit cell $\bm{m}$, induced by an external electric field $\bm E$, is calculated within the linear-response regime, 
\vspace{-0.1cm}
 \begin{equation}
     \bm{m} = (\chi^s + \chi^l) \bm{E} \ .
 \end{equation}

The spin and orbital Edelstein susceptibility tensors, $\chi^s$ and $\chi^l$, are calculated using the semiclassical Boltzmann approach, assuming zero temperature, as

\begin{equation}
    \chi^{\mathcal{O}}_{ij} = \frac{g_{\mathcal{O}} V_0 e \mu_\mathrm{B}}{V_s \hbar} \sum_{\bk} \Braket{\mathcal{O}}^i_{\bk}  \Lambda^j_{\bk} \delta(\energy_{\bk} - \fermi) \ ,
    \label{eq:Edelstein_susceptibility}
\end{equation}

\noindent with $\Braket{\mathcal{O}}^i_{\bk}$ the $i$th component of the expectation value of the spin and orbital angular momentum operators, respectively.
Here, $g_{s(l)}$ indicates the spin(orbital) Landé $g$-factor, $V_0(V_s)$ is the volume of the unit cell (of the system), $\mu_{\text B}$, $\hbar$, $\hbar \bk$, $\energy_{\bk}$, and $\fermi$ represent the Bohr magneton, the reduced Planck constant, the crystal momentum, the energy of the state $\bk$, and the Fermi energy, respectively. Here and in the following, $\bm k$ is a multi-index, including the band index as well as the crystal momentum. Within the constant relaxation time approximation, the mean free path is calculated as $\bm{\Lambda_k} = \tau_0 \bm{v_k}$, with $\tau_0$ the relaxation time and $\bm{v_k} = \frac{1}{\hbar} \frac{\partial \energy_{\bk}}{\partial \bk}$ the group velocity. Using this approximation, the Edelstein susceptibility tensors given by Eq.~\eqref{eq:Edelstein_susceptibility} scale linearly with  $\tau_0$. The Dirac delta function $\delta$ in Eq.~\eqref{eq:Edelstein_susceptibility} indicates that only states at the Fermi surfaces contribute to the nonequilibrium magnetic moment, i.e., states with band energy $\energy_{\bk}$ equal to the Fermi energy $\fermi$. The intrinsic contributions to the charge-spin and charge-orbital conversions vanish due to time-reversal symmetry. Details of the Boltzmann approach and the operators are presented in the Appendices \ref{app:Boltzmann} and \ref{app:Operators}.

\section{Electronic structure}\label{sec:elec_str}

\begin{figure*}
    \includegraphics[width=1.0\linewidth]{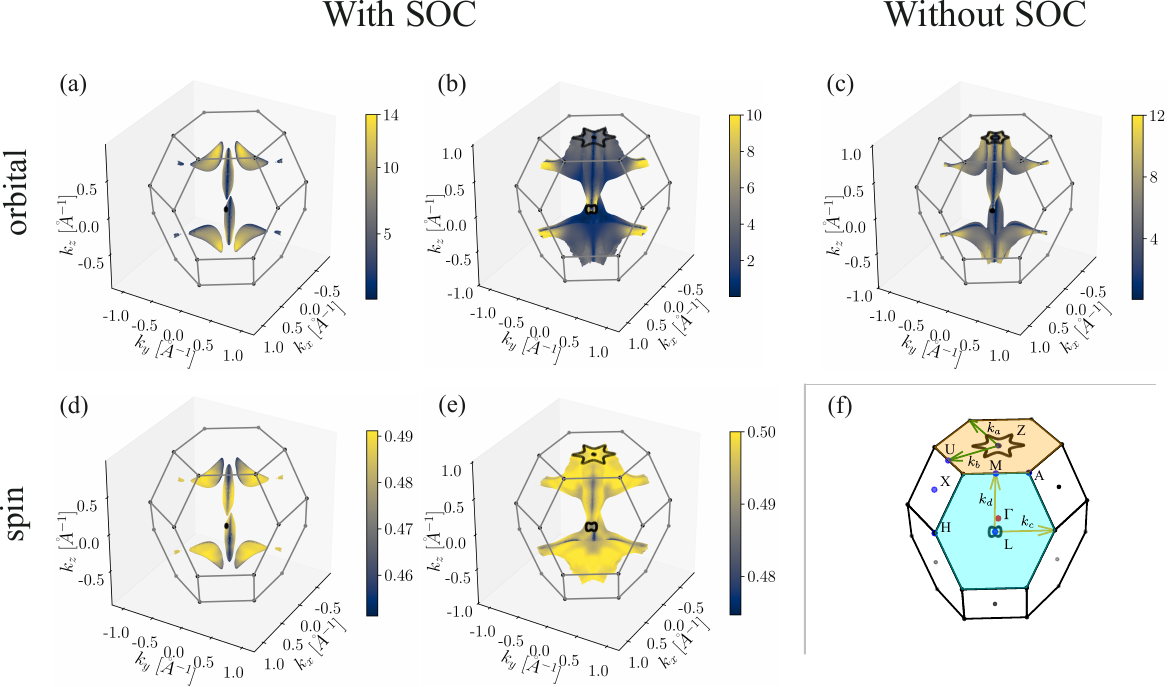}
    \caption{Iso-energy surfaces of the valence band(s) at $\energy = \SI{-0.41}{\electronvolt}$ with respect to $\fermi$. (a) Lower and (b) upper bands of the topmost valence band pair \textcircled{\raisebox{-0.9pt}{$1$}}, with SOC. (c) Valence band without SOC. (a-c) Color shows the absolute value of the OAM in units of $\hbar$. (d-e) Same as (a-b) but with color showing the absolute value of the spin momentum in units of $\hbar$. (f) Brillouin zone with highlighted Z (orange) and L (light blue) planes and relative coordinates $k_a$, $k_b$, $k_c$, $k_d$.}
    \label{fig:surfaces}
\end{figure*}

\begin{figure*}
    \centering
    \includegraphics[width=0.95\linewidth]{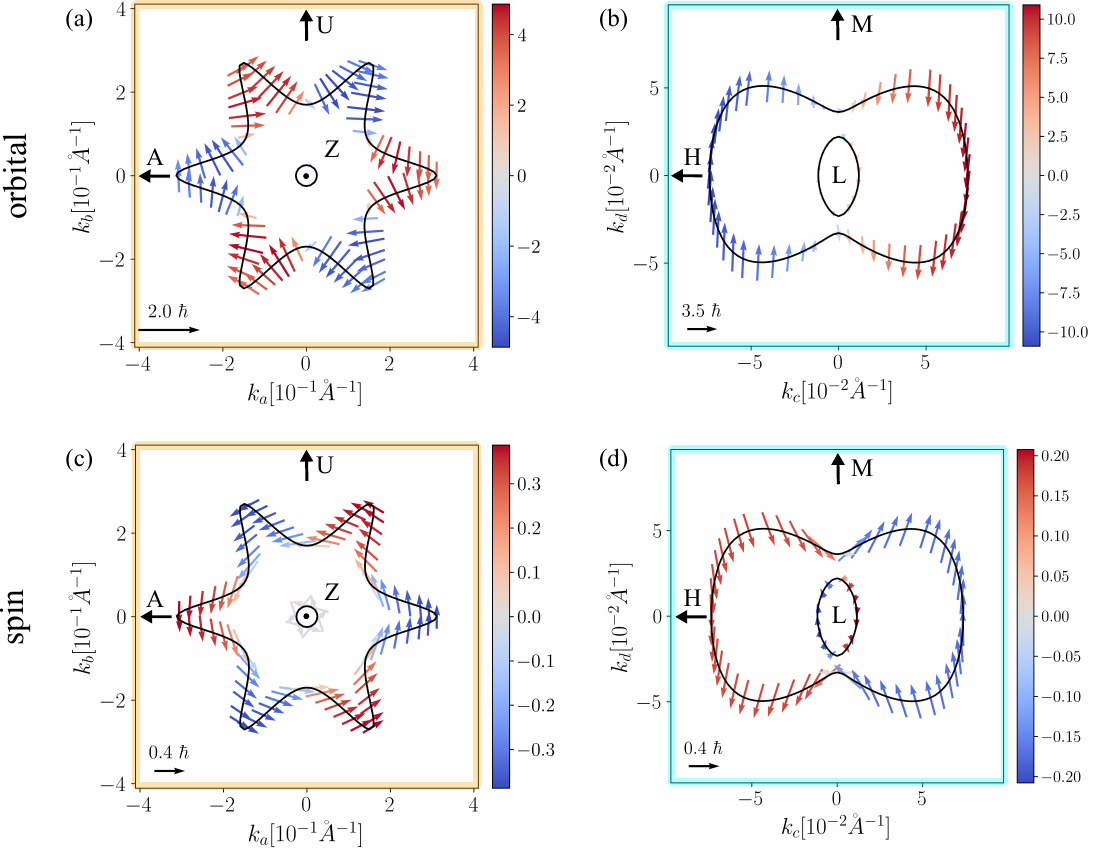}
    \caption{Iso-energy lines at planes $Z$ and $L$ for $\energy =\SI{-0.41}{\electronvolt}$ from the Fermi energy, with the orbital (a-b) and spin (c-d) textures. The lines correspond to the iso-energy $k$-points at the surface of the Brillouin zone, which contain the $Z$ (a,c) and $L$ (b,d) points, highlighted in Fig.~\ref{fig:surfaces}f. The color corresponds to the out-of-plane spin and orbital component in units of $\hbar$, and the arrows indicate the in-plane spin and orbital components in the respective planes. Note that panels~(a,c) and (b,d) use different $k$ scales for better visualization of the iso-energy lines at planes $Z$ and $L$.}
    \label{fig:spin_orbital_texture}
\end{figure*}

Figures \ref{fig:electronic_structure}a-c show the crystal structure of bulk $\alpha$-GeTe, the first Brillouin zone, and the band structure with and without SOC calculated from the Wannier tight-binding model based on the DFT results. The reference energy is set at Fermi energy computed from DFT, the upper valence band energies at $\Gamma$, $L$, and $Z$ are $\SI{-1.57}{\electronvolt}$, $\SI{-0.39}{\electronvolt}$, $\SI{-0.49}{\electronvolt}$, respectively. The relative displacement between the Ge and Te ions along the $[111]$ direction breaks inversion symmetry, introducing Rashba-like splittings, as demonstrated in Refs.~\cite{Ponet2018, DiSante_2013}. Due to this bulk inversion asymmetry, the eight hexagonal faces of the Brillouin zone can be separated into two groups. The first group, with the two faces perpendicular to the $[111]$ direction, has the $Z$ point at their centers, while the second group, the remaining six faces, has the $L$ points at their centers. For simplicity, we refer to the faces orthogonal to the $[111]$ and $[100]$ directions as plane $Z$ and plane $L$, respectively, where plane $Z$ ($L$) contains the $Z$ ($L$) point, as shown in Fig.~\ref{fig:electronic_structure}b. It is important to note that the global valence band maximum and the conduction band minimum are not captured by the high symmetry path in Fig.~\ref{fig:electronic_structure}c.
Around $Z$ and $L$, Rashba-like spin splittings are visible close to the Fermi energy, as shown in detail in Figs.~\ref{fig:electronic_structure}d and e. 
The spin splitting is, however, present only with SOC, and results from the displacement of the Ge atom along $[111]$ leading to the ferroelectric character. Rashba-like splittings also occur around $\Gamma$, located deeper below the Fermi energy.

Figures \ref{fig:surfaces}a-e show the iso-energy surfaces for GeTe with and without SOC at $\SI{-0.41}{\electronvolt}$ relative to the Fermi energy, clearly illustrating the distinction between the two sets of hexagonal faces in the Brillouin zone. As expected from the band structure in Figs.~\ref{fig:electronic_structure}c-e, the iso-energy surfaces in Figs.~\ref{fig:surfaces}b and e show Rashba-like splitting around the $Z$ and $L$ points. Notably, the star-like shape of the iso-energy line around $Z$ and the distinction between $Z$ and $L$, as shown in Fig.~\ref{fig:surfaces}c, are present without SOC.

Based on the energy and momentum splitting in Figs.~\ref{fig:electronic_structure}e and the iso-energy lines in Fig.~\ref{fig:spin_orbital_texture}, the electronic structure of the two topmost valence bands in plane $Z$ can be approximated by an effective Rashba model with warping, as usually occurring in systems with  $C_{3v}$ symmetry~\cite{Fu2009, Johansson2016}, $H_\text{R} = \frac{\hbar^2}{2 m_\text{eff}} k^2 + \alpha_\text{R} (k_a \sigma_y - k_b \sigma_x) + \frac{\lambda}{2} (k^3_+ + k^3_-) \sigma_z$, where $k_\pm = k_a \pm k_b$. Here, $\alpha_\text{R}$ is the effective Rashba parameter, and $\lambda$ is the warping parameter, $k_a$ and $k_b$ are in the $Z$ plane, as indicated in Fig.~\ref{fig:surfaces}f. 

Focusing on the lines around the $Z$-point and neglecting the cubic terms, we use an isotropic Rashba model to estimate characteristic parameters of the valence band in the $\overline{ZA}$ ($\overline{ZU}$) direction as shown in Fig.~\ref{fig:electronic_structure}e: the Rashba energy splitting ~$E_R = \SI{0.310}{\electronvolt} (\SI{0.264}{\electronvolt})$, momentum splitting ~$k_R = 0.123 \text{\AA}^{-1} (0.098 \text{\AA}^{-1})$, effective mass ~$m_\text{eff} = -0.187 m_\text{e} (-0.139 m_\text{e})$ with $m_e$ the electron's mass, and the Rashba parameter ~$\alpha_R = 5.02 \text{eV\AA} (5.37 \text{eV \AA})$. These results are in agreement with previous calculations in Ref.~\cite{DiSante_2013, Picozzi2014, Krempasky2016, Krempasky2020}.

\section{Spin and orbital Edelstein effects}\label{sec:spin_orbital}
Figures \ref{fig:surfaces}a-e show the absolute value of the orbital and spin angular momentum, where spin expectation values are almost constant on the iso-energy surfaces in Fig.~\ref{fig:surfaces}d-e. However, from Figs.~\ref{fig:surfaces}a-c, it is clear that the orbital texture presents hotspots at $L$ faces, which are also visible in Fig.~\ref{fig:spin_orbital_texture}b. Within the ACA, the electrons have mainly $p$ character. Hence, the large values of the orbital angular momentum around $L$ originate from interband and itinerant contributions, which are included in the MTOM. However, the majority of electronic states exhibit orbital angular momenta around $\SIrange{2}{4} \hbar$, indicating that itinerant contributions to the OAM are also relevant here. Similarly large values of the OAM have been predicted for gapped bilayer graphene~\cite{cysne2024controlling}.

Figure \ref{fig:spin_orbital_texture} displays the iso-energy lines on planes $Z$ and $L$ with the expectation values of orbital and spin angular momentum. The coordinate system for each plane is illustrated in Fig.~\ref{fig:surfaces}f, where in plane $Z$, $-k_a$ and $k_b$ are along $\overline{ZA}$ and $\overline{ZU}$ respectively, and analogously $-k_c$ and $k_d$ are along $\overline{LH}$ and $\overline{LM}$ in plane $L$, respectively. Here, the displacement between Ge and Te, combined with SOC, leads to spin splitting of the top valence bands. A comparison of the orbital and spin textures around the $Z$ point in Figs.~\ref{fig:spin_orbital_texture}a and c reveals that in the $Z$ plane, the orbital and spin textures exhibit opposite orientation, with the expectation values of orbital angular momentum being up to $10$ times larger than the spin expectation values. Figure \ref{fig:spin_orbital_texture}c shows good agreement with previous results reported for GeTe \cite{DiSante_2013, Picozzi2014, Tenzin2023}, both in the shape of the iso-energy lines and the spin texture. The spin texture in Fig.~\ref{fig:spin_orbital_texture}c is Rashba-like with approximately perpendicular spin-momentum locking of the in-plane components, with the amplitude almost constant along the iso-energy line, whereas the orbital texture in Fig.~
\ref{fig:spin_orbital_texture}a is weakly unlocked~\cite{Hakioglu2023} with relevant variations in the length of the orbital momenta. 
These variations are most noticeable along the $\overline{ZU}$ direction, where the orbital momentum is minimum. Note that $\mathcal E = - \SI{0.41}{\electronvolt}$ is above the band degeneracy point and hence both iso-energy lines exhibit the same orientation of spin expectation values.

Similarly, the orbital and spin textures along the iso-energy lines around the $L$ point, shown in Fig.~\ref{fig:spin_orbital_texture}b and \ref{fig:spin_orbital_texture}d, exhibit opposite orientation, and a larger orbital than spin momentum. However, in the $L$ plane, the magnitudes of spin and orbital momenta differ by more than one order of magnitude. Remarkably, moving along the iso-energy lines, the in-plane spin texture in Fig.~\ref{fig:spin_orbital_texture}d rotates in the opposite direction to that in Fig.~\ref{fig:spin_orbital_texture}c, clearly deviating from a pure Rashba-like spin texture. On the other hand, the in-plane orbital momentum in Fig.~
\ref{fig:spin_orbital_texture}b is mainly aligned along one direction, and decays rapidly around the $\overline{LM}$ direction. The Rashba-like spin texture in the $Z$ plane follows from the polar displacement of Ge and Te atoms along the $[111]$ direction, whereas in the $L$ plane the spin splitting is not purely Rashba-like due to the different orientation of that plane (perpendicular to $[100]$ and hence not perpendicular to the polarization direction). In Ref.~\cite{Tenzin2023}, the spin texture around $L$ is classified as Rashba-Dresselhaus spin texture. It is important to note that the orbital momenta of both bands share the same sense of rotation. 
In contrast, the spin momenta of the two bands point in opposite directions. Note that $\mathcal E = - \SI{0.41}{\electronvolt}$ is here below the band degeneracy point.

\begin{figure}
    \centering
    \includegraphics[width=1.0\linewidth]{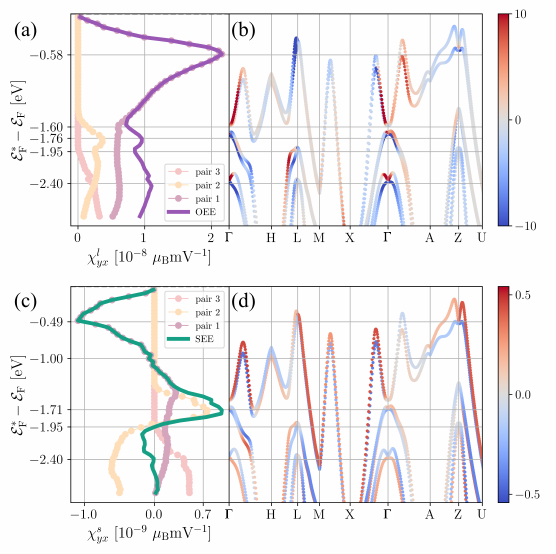}
    \caption{Orbital and spin Edelstein susceptibility. (a) Band-resolved orbital Edelstein susceptibility in GeTe calculated with the Wannier tight-binding model. (b) Band structure with $y$ component of the orbital expectation values in units of $\hbar$. (c), (d) same as (a), (b), but for the spin degree of freedom. Calculations were performed for $V_0 = 56.8 \text{\AA}^3$ , $g_s = 2$, $g_l =1$, and $\tau_0 = \SI{10}{\femto\second}$, which is the typical order of magnitude reported for GeTe~\cite{Kadlec2011,Boschker2018, Guo2019}.}
    \label{fig:SEE_OEE}
\end{figure}

Previous works have shown that switching the ferroelectric polarization along the $\hat{z}$ direction, either by changing the termination \cite{Rinaldi_2016, vojacek_domain_2024} or by an applied electric field \cite{Krempasky2018}, allows tuning the Rashba effect, leading to a reversal of the spin texture and, consequently, the direction of charge-spin conversion~\cite{DiSante_2013, Picozzi2014}. This behavior arises naturally for our bulk system because inverting the ferroelectric polarization is equivalent to maintaining the polarization and inverting the coordinate system. Therefore, it is clear that switching the ferroelectric polarization in our system leads to a reversal of the spin and orbital textures, and hence the charge-spin and charge-orbital conversion.

 Using Eq.~\eqref{eq:Edelstein_susceptibility}, we compute the spin and orbital susceptibility tensors, $\chi^s_{ij}$ and $\chi^l_{ij}$. We simulate a modified Fermi energy $\fermi^*$, focusing on a region below the DFT-calculated Fermi energy $\fermi$ since a p-type doping is observed experimentally in GeTe~\cite{Varotto_2021,Krempasky2016,Yang_2021}, possibly caused by a Te-rich stoichiometry~\cite{Deringer_2012,vojacek_domain_2024}. Due to the $C_{3v}$ symmetry around the $[111]$ direction that defines the $\hat{z}$ direction, the only nonzero components are $\chi^{s/l}_{xy} = -\chi^{s/l}_{yx}$, where the $\hat{x}$ and $\hat{y}$ are defined by $[1 \overline{1} 0]$ and $[1 1 \overline{2}]$, respectively \cite{Johansson2024, Nikolaev2024}. Figure~\ref{fig:SEE_OEE} shows the spin and orbital susceptibilities for energies below the real Fermi energy, as well as the band structure with the $y$ components of spin and orbital expectation values. As expected from the relation between the textures around the $Z$ point, the spin and orbital susceptibilities have opposite signs in the energy range below the valence band maximum; however, the orbital tensor is approximately one order of magnitude larger than the spin, which results from the strongly enhanced orbital expectation values around the $L$ point, as shown in Figs.~~\ref{fig:spin_orbital_texture}b and d. As shown in Appendix \ref{app:maxima_energy}, the global maximum of the valence band is not located along the high-symmetry path and hence not shown in the band structure in Figs.~\ref{fig:electronic_structure}c, \ref{fig:SEE_OEE}b, and \ref{fig:SEE_OEE}d. Nevertheless, these states contribute to the SEE and OEE, and therefore the spin and orbital Edelstein susceptibilities are nonzero already above the local maxima around $Z$, which are visible in the presented band structure.
 
 Figure \ref{fig:SEE_OEE}c shows further the contribution of each band pair to the spin Edelstein susceptibility. Similar to the two-dimensional Rashba model, usually, both bands of a spin-split pair contribute with opposite sign to the spin Edelstein effect, and hence their contributions partially compensate, leading to a reduced total Edelstein response. Sign changes of the total spin Edelstein susceptibility arise from the multiband character of the band structure, which is far more intricate than the two-dimensional effective Rashba model. Hence, the exact energies at which the SEE reverses sign are not directly tied to specific features of the band structure but arise from the multiband character. Comparing Fig.~\ref{fig:SEE_OEE}c with the spin expectation values and the band structure shown in Fig.~\ref{fig:SEE_OEE}d, it becomes obvious that the extrema of the SEE can be related to the pronounced Rashba-like splittings at the $Z$ and $\Gamma$ points, occurring at approximately $\SI{-0.5}{\electronvolt}$ and $\SI{-1.7}{\electronvolt}$, respectively. Remarkably, modifying the Fermi energy $\fermi^*$, e.g. by doping, allows to switch the orientation of the current-induced spin polarization (SEE), without switching the ferroelectric polarization.

 In contrast to the SEE, the OEE does not exhibit any sign changes within the studied energy range, as shown in Fig.~\ref{fig:SEE_OEE}a. Due to the shared sense of rotation, both bands of a pair contribute with the same sign to the orbital Edelstein effect, leading to an enhanced OEE. The orbital Edelstein susceptibility is maximum slightly below the band crossing at the $Z$ point. This maximum, hence, does not result from compensations from the inner band as for the SEE, but from generally smaller orbital expectation values below the band crossing. We observe that the other extrema of the OEE are also not tied to the band crossings, but rather occur slightly below, suggesting a more intricate dependence on the orbital texture of the iso-energy surface. 

Our results for the SEE are consistent with previous observations of spin-charge conversion through spin pumping experiments \cite{Rinaldi_2016, Varotto_2021}, and in particular, the connection between the sign change of the conversion signal~\cite{Varotto_2021} under electrically controllable ferroelectric switching~\cite{Krempasky2018}. In Ref.~\cite{Varotto_2021}, the experimentally observed spin-charge conversion has been found to mainly originate from the inverse spin Hall effect. On the other hand, in Ref.~\cite{Rinaldi_2016} both inverse SEE and inverse spin Hall effect are considered as origin of the spin-charge conversion. Although no quantitative numbers for the spin-charge conversion efficiencies are given, our theoretical results are in agreement with the experimentally observed sign of spin-charge conversion.  However, in Ref.~\cite{Rinaldi_2016} the observed spin-charge conversion could originate from surface as well as bulk states, whereas we consider only the bulk contribution to the SEE.  The spin texture around both studied symmetry points $Z$ and $L$, as well as the SEE in Fig.~\ref{fig:SEE_OEE}c, are in qualitative agreement with the theoretical results in Ref.~\cite{Tenzin2023}. Other ferroelectric materials have been shown to present similar charge-spin conversion efficiencies, such as In$_2$Se$_3$ \cite{Tenzin2023}, WTe$_2$ \cite{jafari2022ferroelectric}, and SnTe~\cite{souza2024unveiling}. Additionally, we find a remarkably large OEE that dominates the direct Edelstein effect in GeTe across a wide energy range, which has not yet been reported either theoretically or experimentally. Therefore, GeTe is not only a very promising material for spintronics and ferroelectric applications but also for orbitronics.

\begin{figure}
    \centering
    \includegraphics[width=1.0\linewidth]{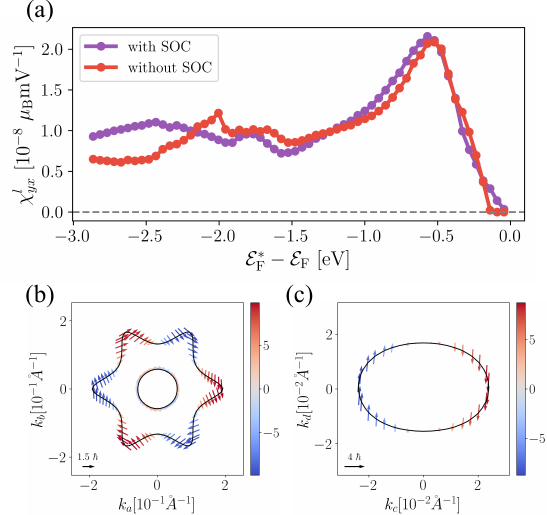}
    \caption{(a) Comparison of the OEE with and without SOC. Iso-energy line and orbital texture of GeTe without SOC on (b) plane $Z$ and (c) plane $L$, at $\SI{-0.41}{\electronvolt}$ relative to the Fermi energy. The arrows represent the orientation of the in-plane components, and the color shows the value of the out-of-plane component in units of $\hbar$. Note the different $k$ scales in (b) and (c) for better visualization of the iso-energy lines.}
    \label{fig:without_SOC}
\end{figure}

In the absence of SOC, the spin Edelstein response naturally vanishes, but an orbital response is expected to exist regardless of the SOC \cite{Go2018}. In Fig.~\ref{fig:without_SOC}a, we compare the orbital Edelstein susceptibilities with and without SOC. The OEE remains largely unchanged with few deviations due to the modified band structure. The approximately constant maxima strongly suggest that the OEE originates mostly from the bulk inversion asymmetry of the crystal structure. Figure \ref{fig:without_SOC}b shows the orbital texture in plane $Z$ without SOC. In agreement with the orbital Rashba effect reported in Ref.~\cite{Ponet2018}, this texture is qualitatively similar to the orbital texture around $Z$ with SOC, shown in Fig.~\ref{fig:spin_orbital_texture}a, but the out-of-plane component is strongly increased. In Fig.~\ref{fig:without_SOC}c, the orbital texture in the $L$ plane is shown, which resembles the orbital texture with SOC. Differences in the size of the iso-energy surfaces and the density of state result from the modified band structure, leading to slightly different transport properties like the orbital Edelstein susceptibility, as shown in Fig.~\ref{fig:without_SOC}a.

\section{Conclusion}
In this work, we have investigated the current-induced spin and orbital magnetization in the bulk ferroelectric Rashba semiconductor GeTe, revealing that the orbital Edelstein effect is approximately one order of magnitude larger than its spin counterpart. The interplay between spin-orbit coupling and the ferroelectric polarization along the [111] direction in GeTe leads to an anisotropic Rashba-like deformation of the Fermi surface, its projection to the $Z$ plane resembling the star-like shape from a warping Rashba model, whereas the Fermi surface projection to the $L$ plane resembles an anisotropic Rashba-Dresselhaus model. Remarkably, the spin Edelstein effect presents clear sign changes within an energy window of approximately $\SI{2}{\electronvolt}$ below the valence band maximum. These sign changes are not linked to ferroelectric switching. Furthermore, we demonstrate that the OEE persists even in the absence of SOC with minimal differences for the top valence bands.

\begin{acknowledgments}
The authors thank Salvatore Teresi and Börge Göbel for insightful discussions throughout the development of this project. This project has received funding from the \textit{European Union's Horizon 2020 research and innovation programme} under grant agreement No~\textit{800945} — NUMERICS — H2020-MSCA-COFUND-2017, and from the \textit{European Union’s Horizon 2020 research and innovation programme} under the Marie Skłodowska-Curie Grant Agreement No~\textit{955671} - SPEAR - H2020-MSCA-COFUND-2020. Additional support was provided by a France 2030 government grant managed by the French National Research Agency PEPR SPIN ANR-22-EXSP 0009 (SPINTHEORY).
\end{acknowledgments}

\section*{Data availability}
The data that support the findings of this article are openly available \cite{leiva2025data}

\appendix
\section{Semiclassical Boltzmann transport theory} \label{app:Boltzmann}

The magnetic moment per unit cell from the spin and orbital angular momentum is given by

\begin{equation}
    \vec{m} = - \frac{\mu_B}{\hbar} \frac{V_0}{V_s} \sum_{n\bk} f_{n\bk} (g_s \vec{s}_{n\bk} + g_l \vec{l}_{n\bk}) \ ,
    \label{eq:m_app}
\end{equation}

\noindent within the semiclassical Boltzmann transport theory. The different factors in Eq.~\eqref{eq:m_app} are explained in Section \ref{sec:methods}. The distribution function $f_{n\bk}$ can be separated in two parts: the equilibrium part $f^0_{n\bk}$, in this case the Fermi-Dirac distribution, and the nonequilibrium part $g_{n\bk}$. It is easily concluded by symmetry analysis that in nonmagnetic materials, the equilibrium contribution would yield no magnetic moment regardless of the source.

For a stationary and spatially homogeneous system, the Boltzmann equation is reduced to 

\begin{equation}
    \dot{\bk} \frac{\partial f_{n\bk}}{\partial \bk} = \left( \frac{\partial f_{n\bk}}{\partial t} \right)_{\mathrm{scatt}}. 
\end{equation}

We can further simplify the above equation if we apply the semiclassical equation of motion in the presence of an electric field, $\dot{\bk} = - \frac{e}{\hbar} \vec{E}$, neglecting scattering-in contributions, and using the linear ansatz for the nonequilibrium part of the distribution function, $g_{n\bk} = \frac{\partial f^0_{n\bk}}{\partial \energy} e \vec{\Lambda}_{n\bk}\cdot \vec{E}$, with $\vec{\Lambda}_{n\bk}$ the mean-free path, so then the Boltzmann equation is linearized and we can solve it for $\vec{\Lambda}_{n\bk}$ as 

\begin{equation}
    \vec{\Lambda}_{n\bk} = \tau_{n\bk} \vec{v}_{n\bk} \ , 
\end{equation}

\noindent with $\tau_{n\bk}$ the momentum relaxation time given by the microscopic transition probability rate $P_{n' \bk' \leftarrow n \bk}$ as,

\begin{equation}
    \frac{1}{\tau_{n\bk}} = \sum_{n' \bk'} P_{(n' \bk') \leftarrow (n \bk)} \ . 
\end{equation}
For the purpose of this paper, we consider a constant relation time $\tau_{n\bk} = \tau_0$ ~\cite{LeivaMontecinos2023, Johansson2024}. 

\section{Spin and orbital angular momentum operators}\label{app:Operators}

The spin operator from the \textit{ab initio} calculations is expressed in the real-space Wannier tight-binding basis $\hat{S}_{\vec{R}}$~\cite{marzari_maximally_1997, marzari_maximally_2012, mostofi_updated_2014} and subsequently used to obtain the interpolated spin expectation values in the reciprocal space $\hat{S}_{\vec{k}}^\mathrm{(H)}$. Our treatment~\cite{vojacek_field-free_2024, vojacek_multiscale_2024} agrees well with the results obtained using the \texttt{wannierBerri} package~\cite{Tsirkin2021}. The orbital angular momentum is computed from the MTOM \cite{thonhauser2011theory} expression 

\begin{equation}\label{eq:l_modern}
    l_{\gamma, n\bk} = \frac{-e}{2 \mu_{\mathrm{B}} g_l} \epsilon_{\alpha \beta \gamma} \text{Im}\braket{\partial_\alpha u_{n\boldsymbol{k}} |\hat{H} - \energy_{n\boldsymbol{k}}| \partial_\beta u_{n\boldsymbol{k}} } \ ,
\end{equation}
with $\Ket{u_{n \bm k}}$ the lattice-periodic part of the Bloch state and $\hat{H}$ the Hamiltonian of the system. Equation~\eqref{eq:l_modern} is directly obtained from the \texttt{wannierBerri} package~\cite{Tsirkin2021}. 

\section{Valence band maximum}
\label{app:maxima_energy}

\begin{figure}
    \centering
    \includegraphics[width=0.9\linewidth]{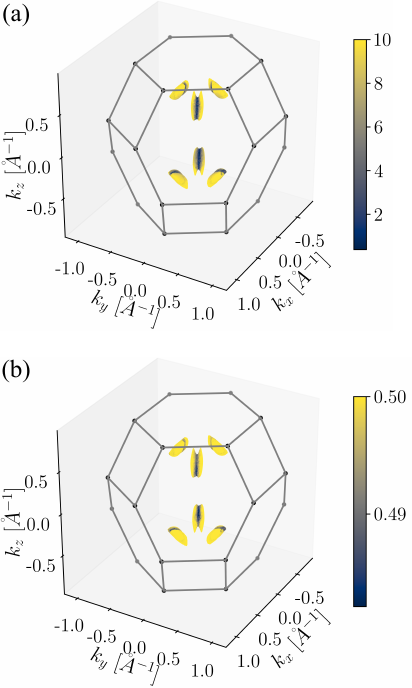}
    \caption{Iso-energy surface of the topmost band close to the global maxima, at $\energy - \fermi = \SI{-0.09}{\electronvolt}$. Color shows the absolute value of the (a) orbital and (b) spin angular momentum, in units of $\hbar$.}
    \label{fig:surface_higher_energy}
\end{figure}

As mentioned above, the global maximum of the valence band at $\energy -\fermi = \SI{-0.05}{\electronvolt}$, is not along the high-symmetry path of the band structure shown in Fig.~\ref{fig:electronic_structure}. Figure \ref{fig:surface_higher_energy} shows the iso-energy surface at $\energy -\fermi = \SI{-0.09}{\electronvolt}$, with the absolute value of the orbital and spin angular momentum. 
For bulk states in GeTe, it has been shown that the maximum of the valence band is very close to the Fermi energy \cite{Kremer2020, Krempasky2020}. Interestingly, as shown in Fig.~\ref{fig:SEE_OEE}, the SEE and OEE are still present this close to the Fermi energy, with a similar absolute value of the spin and orbital angular momenta to Fig.~\ref{fig:surfaces}. This indicates that the decrease in spin and orbital susceptibilities in Fig.~\ref{fig:SEE_OEE} comes from the decreasing size of the iso-energy surface, rather than from a decrease of the spin and orbital momenta. 

\bibliography{Bibliography}

\begin{thebibliography}{94}%
\makeatletter
\providecommand \@ifxundefined [1]{%
 \@ifx{#1\undefined}
}%
\providecommand \@ifnum [1]{%
 \ifnum #1\expandafter \@firstoftwo
 \else \expandafter \@secondoftwo
 \fi
}%
\providecommand \@ifx [1]{%
 \ifx #1\expandafter \@firstoftwo
 \else \expandafter \@secondoftwo
 \fi
}%
\providecommand \natexlab [1]{#1}%
\providecommand \enquote  [1]{``#1''}%
\providecommand \bibnamefont  [1]{#1}%
\providecommand \bibfnamefont [1]{#1}%
\providecommand \citenamefont [1]{#1}%
\providecommand \href@noop [0]{\@secondoftwo}%
\providecommand \href [0]{\begingroup \@sanitize@url \@href}%
\providecommand \@href[1]{\@@startlink{#1}\@@href}%
\providecommand \@@href[1]{\endgroup#1\@@endlink}%
\providecommand \@sanitize@url [0]{\catcode `\\12\catcode `\$12\catcode `\&12\catcode `\#12\catcode `\^12\catcode `\_12\catcode `\%12\relax}%
\providecommand \@@startlink[1]{}%
\providecommand \@@endlink[0]{}%
\providecommand \url  [0]{\begingroup\@sanitize@url \@url }%
\providecommand \@url [1]{\endgroup\@href {#1}{\urlprefix }}%
\providecommand \urlprefix  [0]{URL }%
\providecommand \Eprint [0]{\href }%
\providecommand \doibase [0]{https://doi.org/}%
\providecommand \selectlanguage [0]{\@gobble}%
\providecommand \bibinfo  [0]{\@secondoftwo}%
\providecommand \bibfield  [0]{\@secondoftwo}%
\providecommand \translation [1]{[#1]}%
\providecommand \BibitemOpen [0]{}%
\providecommand \bibitemStop [0]{}%
\providecommand \bibitemNoStop [0]{.\EOS\space}%
\providecommand \EOS [0]{\spacefactor3000\relax}%
\providecommand \BibitemShut  [1]{\csname bibitem#1\endcsname}%
\let\auto@bib@innerbib\@empty
\bibitem [{\citenamefont {Rashba}\ and\ \citenamefont {Sheka}(1959)}]{rashba2015symmetry}%
  \BibitemOpen
  \bibfield  {author} {\bibinfo {author} {\bibfnamefont {E.}~\bibnamefont {Rashba}}\ and\ \bibinfo {author} {\bibfnamefont {V.~I.}\ \bibnamefont {Sheka}},\ }\bibfield  {title} {\bibinfo {title} {{Symmetry of energy bands in crystals of wurtzite type II. Symmetry of bands with spin-orbit interaction included}},\ }\href@noop {} {\bibfield  {journal} {\bibinfo  {journal} {Fizika Tverdogo Tela: Collected Papers}\ }\textbf {\bibinfo {volume} {2}},\ \bibinfo {pages} {162} (\bibinfo {year} {1959})},\ \bibinfo {note} {in Russian}\BibitemShut {NoStop}%
\bibitem [{\citenamefont {Bihlmayer}\ \emph {et~al.}(2015)\citenamefont {Bihlmayer}, \citenamefont {Rader},\ and\ \citenamefont {Winkler}}]{Bihlmayer2015}%
  \BibitemOpen
  \bibfield  {author} {\bibinfo {author} {\bibfnamefont {G.}~\bibnamefont {Bihlmayer}}, \bibinfo {author} {\bibfnamefont {O.}~\bibnamefont {Rader}},\ and\ \bibinfo {author} {\bibfnamefont {R.}~\bibnamefont {Winkler}},\ }\bibfield  {title} {\bibinfo {title} {{Focus on the Rashba effect}},\ }\href {https://doi.org/10.1088/1367-2630/17/5/050202} {\bibfield  {journal} {\bibinfo  {journal} {New Journal of Physics}\ }\textbf {\bibinfo {volume} {17}},\ \bibinfo {pages} {050202} (\bibinfo {year} {2015})},\ \bibinfo {note} {{English translation of Ref.~[1] in the supplementary material.}}\BibitemShut {Stop}%
\bibitem [{\citenamefont {Rashba}(1960)}]{Rashba1960}%
  \BibitemOpen
  \bibfield  {author} {\bibinfo {author} {\bibfnamefont {E.}~\bibnamefont {Rashba}},\ }\bibfield  {title} {\bibinfo {title} {{{Properties of semiconductors with an extremum loop. 1. Cyclotron and combinational resonance in a magnetic field perpendicular to the plane of the loop}}},\ }\href@noop {} {\bibfield  {journal} {\bibinfo  {journal} {Sov. Phys. Solid State}\ }\textbf {\bibinfo {volume} {2}},\ \bibinfo {pages} {1109} (\bibinfo {year} {1960})}\BibitemShut {NoStop}%
\bibitem [{\citenamefont {Bychkov}\ and\ \citenamefont {Rashba}(1984{\natexlab{a}})}]{Rashba1984}%
  \BibitemOpen
  \bibfield  {author} {\bibinfo {author} {\bibfnamefont {Y.}~\bibnamefont {Bychkov}}\ and\ \bibinfo {author} {\bibfnamefont {E.}~\bibnamefont {Rashba}},\ }\bibfield  {title} {\bibinfo {title} {{{Properties of a 2D electron gas with lifted spectral degeneracy}}},\ }\href {http://www.jetpletters.ac.ru/ps/1264/article_19121.pdf} {\bibfield  {journal} {\bibinfo  {journal} {JETP}\ }\textbf {\bibinfo {volume} {39}},\ \bibinfo {pages} {78} (\bibinfo {year} {1984}{\natexlab{a}})}\BibitemShut {NoStop}%
\bibitem [{\citenamefont {Bychkov}\ and\ \citenamefont {Rashba}(1984{\natexlab{b}})}]{Rashba1984_2}%
  \BibitemOpen
  \bibfield  {author} {\bibinfo {author} {\bibfnamefont {Y.~A.}\ \bibnamefont {Bychkov}}\ and\ \bibinfo {author} {\bibfnamefont {E.~I.}\ \bibnamefont {Rashba}},\ }\bibfield  {title} {\bibinfo {title} {{Oscillatory effects and the magnetic susceptibility of carriers in inversion layers}},\ }\href {http://stacks.iop.org/0022-3719/17/i=33/a=015} {\bibfield  {journal} {\bibinfo  {journal} {J. of Phys. C}\ }\textbf {\bibinfo {volume} {17}},\ \bibinfo {pages} {6039} (\bibinfo {year} {1984}{\natexlab{b}})}\BibitemShut {NoStop}%
\bibitem [{\citenamefont {Di~Sante}\ \emph {et~al.}(2013)\citenamefont {Di~Sante}, \citenamefont {Barone}, \citenamefont {Bertacco},\ and\ \citenamefont {Picozzi}}]{DiSante_2013}%
  \BibitemOpen
  \bibfield  {author} {\bibinfo {author} {\bibfnamefont {D.}~\bibnamefont {Di~Sante}}, \bibinfo {author} {\bibfnamefont {P.}~\bibnamefont {Barone}}, \bibinfo {author} {\bibfnamefont {R.}~\bibnamefont {Bertacco}},\ and\ \bibinfo {author} {\bibfnamefont {S.}~\bibnamefont {Picozzi}},\ }\bibfield  {title} {\bibinfo {title} {{Electric Control of the Giant Rashba Effect in Bulk GeTe}},\ }\href {https://doi.org/https://doi.org/10.1002/adma.201203199} {\bibfield  {journal} {\bibinfo  {journal} {Advanced Materials}\ }\textbf {\bibinfo {volume} {25}},\ \bibinfo {pages} {509} (\bibinfo {year} {2013})}\BibitemShut {NoStop}%
\bibitem [{\citenamefont {Picozzi}(2014)}]{Picozzi2014}%
  \BibitemOpen
  \bibfield  {author} {\bibinfo {author} {\bibfnamefont {S.}~\bibnamefont {Picozzi}},\ }\bibfield  {title} {\bibinfo {title} {{Ferroelectric Rashba semiconductors as a novel class of multifunctional materials}},\ }\bibfield  {journal} {\bibinfo  {journal} {Frontiers in Physics}\ }\textbf {\bibinfo {volume} {2}},\ \href {https://doi.org/10.3389/fphy.2014.00010} {10.3389/fphy.2014.00010} (\bibinfo {year} {2014})\BibitemShut {NoStop}%
\bibitem [{\citenamefont {Sharma}\ \emph {et~al.}(2022)\citenamefont {Sharma}, \citenamefont {Moise}, \citenamefont {Colombo},\ and\ \citenamefont {Seidel}}]{Sharma2021}%
  \BibitemOpen
  \bibfield  {author} {\bibinfo {author} {\bibfnamefont {P.}~\bibnamefont {Sharma}}, \bibinfo {author} {\bibfnamefont {T.~S.}\ \bibnamefont {Moise}}, \bibinfo {author} {\bibfnamefont {L.}~\bibnamefont {Colombo}},\ and\ \bibinfo {author} {\bibfnamefont {J.}~\bibnamefont {Seidel}},\ }\bibfield  {title} {\bibinfo {title} {{Roadmap for Ferroelectric Domain Wall Nanoelectronics}},\ }\href {https://doi.org/https://doi.org/10.1002/adfm.202110263} {\bibfield  {journal} {\bibinfo  {journal} {Advanced Functional Materials}\ }\textbf {\bibinfo {volume} {32}},\ \bibinfo {pages} {2110263} (\bibinfo {year} {2022})}\BibitemShut {NoStop}%
\bibitem [{\citenamefont {Gu}\ \emph {et~al.}(2024)\citenamefont {Gu}, \citenamefont {Zheng}, \citenamefont {Jia}, \citenamefont {Shi}, \citenamefont {Zhao}, \citenamefont {Zeng}, \citenamefont {Zhang}, \citenamefont {Zhu}, \citenamefont {Wang},\ and\ \citenamefont {Chen}}]{Gu2024}%
  \BibitemOpen
  \bibfield  {author} {\bibinfo {author} {\bibfnamefont {Y.}~\bibnamefont {Gu}}, \bibinfo {author} {\bibfnamefont {Z.}~\bibnamefont {Zheng}}, \bibinfo {author} {\bibfnamefont {L.}~\bibnamefont {Jia}}, \bibinfo {author} {\bibfnamefont {S.}~\bibnamefont {Shi}}, \bibinfo {author} {\bibfnamefont {T.}~\bibnamefont {Zhao}}, \bibinfo {author} {\bibfnamefont {T.}~\bibnamefont {Zeng}}, \bibinfo {author} {\bibfnamefont {Q.}~\bibnamefont {Zhang}}, \bibinfo {author} {\bibfnamefont {Y.}~\bibnamefont {Zhu}}, \bibinfo {author} {\bibfnamefont {H.}~\bibnamefont {Wang}},\ and\ \bibinfo {author} {\bibfnamefont {J.}~\bibnamefont {Chen}},\ }\bibfield  {title} {\bibinfo {title} {Ferroelectric control of spin-orbitronics},\ }\href {https://doi.org/https://doi.org/10.1002/adfm.202406444} {\bibfield  {journal} {\bibinfo  {journal} {Advanced Functional Materials}\ }\textbf {\bibinfo {volume} {34}},\ \bibinfo {pages} {2406444} (\bibinfo {year} {2024})}\BibitemShut {NoStop}%
\bibitem [{\citenamefont {Rinaldi}\ \emph {et~al.}(2016)\citenamefont {Rinaldi}, \citenamefont {Rojas-Sánchez}, \citenamefont {Wang}, \citenamefont {Fu}, \citenamefont {Oyarzun}, \citenamefont {Vila}, \citenamefont {Bertoli}, \citenamefont {Asa}, \citenamefont {Baldrati}, \citenamefont {Cantoni}, \citenamefont {George}, \citenamefont {Calarco}, \citenamefont {Fert},\ and\ \citenamefont {Bertacco}}]{Rinaldi_2016}%
  \BibitemOpen
  \bibfield  {author} {\bibinfo {author} {\bibfnamefont {C.}~\bibnamefont {Rinaldi}}, \bibinfo {author} {\bibfnamefont {J.~C.}\ \bibnamefont {Rojas-Sánchez}}, \bibinfo {author} {\bibfnamefont {R.~N.}\ \bibnamefont {Wang}}, \bibinfo {author} {\bibfnamefont {Y.}~\bibnamefont {Fu}}, \bibinfo {author} {\bibfnamefont {S.}~\bibnamefont {Oyarzun}}, \bibinfo {author} {\bibfnamefont {L.}~\bibnamefont {Vila}}, \bibinfo {author} {\bibfnamefont {S.}~\bibnamefont {Bertoli}}, \bibinfo {author} {\bibfnamefont {M.}~\bibnamefont {Asa}}, \bibinfo {author} {\bibfnamefont {L.}~\bibnamefont {Baldrati}}, \bibinfo {author} {\bibfnamefont {M.}~\bibnamefont {Cantoni}}, \bibinfo {author} {\bibfnamefont {J.-M.}\ \bibnamefont {George}}, \bibinfo {author} {\bibfnamefont {R.}~\bibnamefont {Calarco}}, \bibinfo {author} {\bibfnamefont {A.}~\bibnamefont {Fert}},\ and\ \bibinfo {author} {\bibfnamefont {R.}~\bibnamefont {Bertacco}},\ }\bibfield  {title} {\bibinfo {title} {{Evidence for spin to charge conversion in GeTe(111)}},\ }\href
  {https://doi.org/10.1063/1.4941276} {\bibfield  {journal} {\bibinfo  {journal} {APL Materials}\ }\textbf {\bibinfo {volume} {4}},\ \bibinfo {pages} {032501} (\bibinfo {year} {2016})}\BibitemShut {NoStop}%
\bibitem [{\citenamefont {Rinaldi}\ \emph {et~al.}(2018)\citenamefont {Rinaldi}, \citenamefont {Varotto}, \citenamefont {Asa}, \citenamefont {Sławińska}, \citenamefont {Fujii}, \citenamefont {Vinai}, \citenamefont {Cecchi}, \citenamefont {Di~Sante}, \citenamefont {Calarco}, \citenamefont {Vobornik}, \citenamefont {Panaccione}, \citenamefont {Picozzi},\ and\ \citenamefont {Bertacco}}]{Rinaldi_2018}%
  \BibitemOpen
  \bibfield  {author} {\bibinfo {author} {\bibfnamefont {C.}~\bibnamefont {Rinaldi}}, \bibinfo {author} {\bibfnamefont {S.}~\bibnamefont {Varotto}}, \bibinfo {author} {\bibfnamefont {M.}~\bibnamefont {Asa}}, \bibinfo {author} {\bibfnamefont {J.}~\bibnamefont {Sławińska}}, \bibinfo {author} {\bibfnamefont {J.}~\bibnamefont {Fujii}}, \bibinfo {author} {\bibfnamefont {G.}~\bibnamefont {Vinai}}, \bibinfo {author} {\bibfnamefont {S.}~\bibnamefont {Cecchi}}, \bibinfo {author} {\bibfnamefont {D.}~\bibnamefont {Di~Sante}}, \bibinfo {author} {\bibfnamefont {R.}~\bibnamefont {Calarco}}, \bibinfo {author} {\bibfnamefont {I.}~\bibnamefont {Vobornik}}, \bibinfo {author} {\bibfnamefont {G.}~\bibnamefont {Panaccione}}, \bibinfo {author} {\bibfnamefont {S.}~\bibnamefont {Picozzi}},\ and\ \bibinfo {author} {\bibfnamefont {R.}~\bibnamefont {Bertacco}},\ }\bibfield  {title} {\bibinfo {title} {{Ferroelectric Control of the Spin Texture in GeTe}},\ }\href {https://doi.org/10.1021/acs.nanolett.7b04829} {\bibfield  {journal}
  {\bibinfo  {journal} {Nano Letters}\ }\textbf {\bibinfo {volume} {18}},\ \bibinfo {pages} {2751} (\bibinfo {year} {2018})}\BibitemShut {NoStop}%
\bibitem [{\citenamefont {Varotto}\ \emph {et~al.}(2021)\citenamefont {Varotto}, \citenamefont {Nessi}, \citenamefont {Cecchi}, \citenamefont {Sławińska}, \citenamefont {Noël}, \citenamefont {Petrò}, \citenamefont {Fagiani}, \citenamefont {Novati}, \citenamefont {Cantoni}, \citenamefont {Petti}, \citenamefont {Albisetti}, \citenamefont {Costa}, \citenamefont {Calarco}, \citenamefont {Buongiorno~Nardelli}, \citenamefont {Bibes}, \citenamefont {Picozzi}, \citenamefont {Attané}, \citenamefont {Vila}, \citenamefont {Bertacco},\ and\ \citenamefont {Rinaldi}}]{Varotto_2021}%
  \BibitemOpen
  \bibfield  {author} {\bibinfo {author} {\bibfnamefont {S.}~\bibnamefont {Varotto}}, \bibinfo {author} {\bibfnamefont {L.}~\bibnamefont {Nessi}}, \bibinfo {author} {\bibfnamefont {S.}~\bibnamefont {Cecchi}}, \bibinfo {author} {\bibfnamefont {J.}~\bibnamefont {Sławińska}}, \bibinfo {author} {\bibfnamefont {P.}~\bibnamefont {Noël}}, \bibinfo {author} {\bibfnamefont {S.}~\bibnamefont {Petrò}}, \bibinfo {author} {\bibfnamefont {F.}~\bibnamefont {Fagiani}}, \bibinfo {author} {\bibfnamefont {A.}~\bibnamefont {Novati}}, \bibinfo {author} {\bibfnamefont {M.}~\bibnamefont {Cantoni}}, \bibinfo {author} {\bibfnamefont {D.}~\bibnamefont {Petti}}, \bibinfo {author} {\bibfnamefont {E.}~\bibnamefont {Albisetti}}, \bibinfo {author} {\bibfnamefont {M.}~\bibnamefont {Costa}}, \bibinfo {author} {\bibfnamefont {R.}~\bibnamefont {Calarco}}, \bibinfo {author} {\bibfnamefont {M.}~\bibnamefont {Buongiorno~Nardelli}}, \bibinfo {author} {\bibfnamefont {M.}~\bibnamefont {Bibes}}, \bibinfo {author} {\bibfnamefont {S.}~\bibnamefont
  {Picozzi}}, \bibinfo {author} {\bibfnamefont {J.-P.}\ \bibnamefont {Attané}}, \bibinfo {author} {\bibfnamefont {L.}~\bibnamefont {Vila}}, \bibinfo {author} {\bibfnamefont {R.}~\bibnamefont {Bertacco}},\ and\ \bibinfo {author} {\bibfnamefont {C.}~\bibnamefont {Rinaldi}},\ }\bibfield  {title} {\bibinfo {title} {{Room-temperature ferroelectric switching of spin-to-charge conversion in germanium telluride}},\ }\href {https://doi.org/10.1038/s41928-021-00653-2} {\bibfield  {journal} {\bibinfo  {journal} {Nature Electronics}\ }\textbf {\bibinfo {volume} {4}},\ \bibinfo {pages} {740–747} (\bibinfo {year} {2021})}\BibitemShut {NoStop}%
\bibitem [{\citenamefont {Vojáček}\ \emph {et~al.}(2024{\natexlab{a}})\citenamefont {Vojáček}, \citenamefont {Chshiev},\ and\ \citenamefont {Li}}]{Vojek2024}%
  \BibitemOpen
  \bibfield  {author} {\bibinfo {author} {\bibfnamefont {L.}~\bibnamefont {Vojáček}}, \bibinfo {author} {\bibfnamefont {M.}~\bibnamefont {Chshiev}},\ and\ \bibinfo {author} {\bibfnamefont {J.}~\bibnamefont {Li}},\ }\bibfield  {title} {\bibinfo {title} {{Domain Wall Migration-Mediated Ferroelectric Switching and Rashba Effect Tuning in GeTe Thin Films}},\ }\href {https://doi.org/10.1021/acsaelm.4c00392} {\bibfield  {journal} {\bibinfo  {journal} {ACS Applied Electronic Materials}\ }\textbf {\bibinfo {volume} {6}},\ \bibinfo {pages} {3754–3758} (\bibinfo {year} {2024}{\natexlab{a}})}\BibitemShut {NoStop}%
\bibitem [{\citenamefont {Aronov}\ and\ \citenamefont {Lyanda-Geller}(1989)}]{aronov1989nuclear}%
  \BibitemOpen
  \bibfield  {author} {\bibinfo {author} {\bibfnamefont {A.~G.}\ \bibnamefont {Aronov}}\ and\ \bibinfo {author} {\bibfnamefont {Y.~B.}\ \bibnamefont {Lyanda-Geller}},\ }\bibfield  {title} {\bibinfo {title} {{Nuclear electric resonance and orientation of carrier spins by an electric field}},\ }\href@noop {} {\bibfield  {journal} {\bibinfo  {journal} {JETP Lett.}\ }\textbf {\bibinfo {volume} {50}},\ \bibinfo {pages} {431} (\bibinfo {year} {1989})}\BibitemShut {NoStop}%
\bibitem [{\citenamefont {Edelstein}(1990)}]{edelstein1990spin}%
  \BibitemOpen
  \bibfield  {author} {\bibinfo {author} {\bibfnamefont {V.~M.}\ \bibnamefont {Edelstein}},\ }\bibfield  {title} {\bibinfo {title} {{Spin polarization of conduction electrons induced by electric current in two-dimensional asymmetric electron systems}},\ }\href@noop {} {\bibfield  {journal} {\bibinfo  {journal} {Solid State Communications}\ }\textbf {\bibinfo {volume} {73}},\ \bibinfo {pages} {233} (\bibinfo {year} {1990})}\BibitemShut {NoStop}%
\bibitem [{\citenamefont {Leiva-Montecinos}\ \emph {et~al.}(2023)\citenamefont {Leiva-Montecinos}, \citenamefont {Henk}, \citenamefont {Mertig},\ and\ \citenamefont {Johansson}}]{LeivaMontecinos2023}%
  \BibitemOpen
  \bibfield  {author} {\bibinfo {author} {\bibfnamefont {S.}~\bibnamefont {Leiva-Montecinos}}, \bibinfo {author} {\bibfnamefont {J.}~\bibnamefont {Henk}}, \bibinfo {author} {\bibfnamefont {I.}~\bibnamefont {Mertig}},\ and\ \bibinfo {author} {\bibfnamefont {A.}~\bibnamefont {Johansson}},\ }\bibfield  {title} {\bibinfo {title} {{Spin and orbital Edelstein effect in a bilayer system with Rashba interaction}},\ }\href {https://doi.org/10.1103/PhysRevResearch.5.043294} {\bibfield  {journal} {\bibinfo  {journal} {Phys. Rev. Res.}\ }\textbf {\bibinfo {volume} {5}},\ \bibinfo {pages} {043294} (\bibinfo {year} {2023})}\BibitemShut {NoStop}%
\bibitem [{\citenamefont {Fontenele}\ \emph {et~al.}(2024)\citenamefont {Fontenele}, \citenamefont {dos Anjos Sousa~Júnior}, \citenamefont {Cysne},\ and\ \citenamefont {Costa}}]{fontenele2024impact}%
  \BibitemOpen
  \bibfield  {author} {\bibinfo {author} {\bibfnamefont {R.~A.}\ \bibnamefont {Fontenele}}, \bibinfo {author} {\bibfnamefont {S.}~\bibnamefont {dos Anjos Sousa~Júnior}}, \bibinfo {author} {\bibfnamefont {T.~P.}\ \bibnamefont {Cysne}},\ and\ \bibinfo {author} {\bibfnamefont {N.~C.}\ \bibnamefont {Costa}},\ }\bibfield  {title} {\bibinfo {title} {{The impact of Rashba spin-orbit coupling in charge-ordered systems}},\ }\href {https://doi.org/10.1088/1361-648X/ad2cc9} {\bibfield  {journal} {\bibinfo  {journal} {Journal of Physics: Condensed Matter}\ }\textbf {\bibinfo {volume} {36}},\ \bibinfo {pages} {225601} (\bibinfo {year} {2024})}\BibitemShut {NoStop}%
\bibitem [{\citenamefont {Johansson}\ \emph {et~al.}(2016)\citenamefont {Johansson}, \citenamefont {Henk},\ and\ \citenamefont {Mertig}}]{Johansson2016}%
  \BibitemOpen
  \bibfield  {author} {\bibinfo {author} {\bibfnamefont {A.}~\bibnamefont {Johansson}}, \bibinfo {author} {\bibfnamefont {J.}~\bibnamefont {Henk}},\ and\ \bibinfo {author} {\bibfnamefont {I.}~\bibnamefont {Mertig}},\ }\bibfield  {title} {\bibinfo {title} {{Theoretical aspects of the Edelstein effect for anisotropic two-dimensional electron gas and topological insulators}},\ }\href {https://doi.org/10.1103/PhysRevB.93.195440} {\bibfield  {journal} {\bibinfo  {journal} {Phys. Rev. B}\ }\textbf {\bibinfo {volume} {93}},\ \bibinfo {pages} {195440} (\bibinfo {year} {2016})}\BibitemShut {NoStop}%
\bibitem [{\citenamefont {Gaiardoni}\ \emph {et~al.}(2025)\citenamefont {Gaiardoni}, \citenamefont {Trama}, \citenamefont {Maiellaro}, \citenamefont {Guarcello}, \citenamefont {Romeo},\ and\ \citenamefont {Citro}}]{Gaiardoni2025}%
  \BibitemOpen
  \bibfield  {author} {\bibinfo {author} {\bibfnamefont {I.}~\bibnamefont {Gaiardoni}}, \bibinfo {author} {\bibfnamefont {M.}~\bibnamefont {Trama}}, \bibinfo {author} {\bibfnamefont {A.}~\bibnamefont {Maiellaro}}, \bibinfo {author} {\bibfnamefont {C.}~\bibnamefont {Guarcello}}, \bibinfo {author} {\bibfnamefont {F.}~\bibnamefont {Romeo}},\ and\ \bibinfo {author} {\bibfnamefont {R.}~\bibnamefont {Citro}},\ }\bibfield  {title} {\bibinfo {title} {{Edelstein Effect in Isotropic and Anisotropic Rashba Models}},\ }\href {https://doi.org/10.3390/condmat10010015} {\bibfield  {journal} {\bibinfo  {journal} {Condensed Matter}\ }\textbf {\bibinfo {volume} {10}},\ \bibinfo {pages} {15} (\bibinfo {year} {2025})}\BibitemShut {NoStop}%
\bibitem [{\citenamefont {Dey}\ \emph {et~al.}(2025)\citenamefont {Dey}, \citenamefont {Nandy},\ and\ \citenamefont {Saha}}]{dey2025current}%
  \BibitemOpen
  \bibfield  {author} {\bibinfo {author} {\bibfnamefont {A.}~\bibnamefont {Dey}}, \bibinfo {author} {\bibfnamefont {A.~K.}\ \bibnamefont {Nandy}},\ and\ \bibinfo {author} {\bibfnamefont {K.}~\bibnamefont {Saha}},\ }\bibfield  {title} {\bibinfo {title} {{Current-induced spin polarisation in Rashba–Dresselhaus systems under different point groups}},\ }\href {https://doi.org/10.1088/1367-2630/adac88} {\bibfield  {journal} {\bibinfo  {journal} {New Journal of Physics}\ }\textbf {\bibinfo {volume} {27}},\ \bibinfo {pages} {013024} (\bibinfo {year} {2025})}\BibitemShut {NoStop}%
\bibitem [{\citenamefont {Culcer}\ \emph {et~al.}(2010)\citenamefont {Culcer}, \citenamefont {Hwang}, \citenamefont {Stanescu},\ and\ \citenamefont {Das~Sarma}}]{Culcer2010}%
  \BibitemOpen
  \bibfield  {author} {\bibinfo {author} {\bibfnamefont {D.}~\bibnamefont {Culcer}}, \bibinfo {author} {\bibfnamefont {E.~H.}\ \bibnamefont {Hwang}}, \bibinfo {author} {\bibfnamefont {T.~D.}\ \bibnamefont {Stanescu}},\ and\ \bibinfo {author} {\bibfnamefont {S.}~\bibnamefont {Das~Sarma}},\ }\bibfield  {title} {\bibinfo {title} {{Two-dimensional surface charge transport in topological insulators}},\ }\href {https://doi.org/10.1103/PhysRevB.82.155457} {\bibfield  {journal} {\bibinfo  {journal} {Phys. Rev. B}\ }\textbf {\bibinfo {volume} {82}},\ \bibinfo {pages} {155457} (\bibinfo {year} {2010})}\BibitemShut {NoStop}%
\bibitem [{\citenamefont {Mellnik}\ \emph {et~al.}(2010)\citenamefont {Mellnik}, \citenamefont {Lee}, \citenamefont {Richardella}, \citenamefont {Grab}, \citenamefont {Mintun}, \citenamefont {Fischer}, \citenamefont {Vaezi}, \citenamefont {Manchon}, \citenamefont {Kim}, \citenamefont {Samarth},\ and\ \citenamefont {Ralph}}]{Mellnik2010}%
  \BibitemOpen
  \bibfield  {author} {\bibinfo {author} {\bibfnamefont {A.~R.}\ \bibnamefont {Mellnik}}, \bibinfo {author} {\bibfnamefont {J.~S.}\ \bibnamefont {Lee}}, \bibinfo {author} {\bibfnamefont {A.}~\bibnamefont {Richardella}}, \bibinfo {author} {\bibfnamefont {J.~L.}\ \bibnamefont {Grab}}, \bibinfo {author} {\bibfnamefont {P.~J.}\ \bibnamefont {Mintun}}, \bibinfo {author} {\bibfnamefont {M.~H.}\ \bibnamefont {Fischer}}, \bibinfo {author} {\bibfnamefont {A.}~\bibnamefont {Vaezi}}, \bibinfo {author} {\bibfnamefont {A.}~\bibnamefont {Manchon}}, \bibinfo {author} {\bibfnamefont {E.-A.}\ \bibnamefont {Kim}}, \bibinfo {author} {\bibfnamefont {N.}~\bibnamefont {Samarth}},\ and\ \bibinfo {author} {\bibfnamefont {D.~C.}\ \bibnamefont {Ralph}},\ }\bibfield  {title} {\bibinfo {title} {{Spin-transfer torque generated by a topological insulator}},\ }\href {https://doi.org/10.1038/nature13534} {\bibfield  {journal} {\bibinfo  {journal} {Nature}\ }\textbf {\bibinfo {volume} {511}},\ \bibinfo {pages} {449} (\bibinfo {year}
  {2010})}\BibitemShut {NoStop}%
\bibitem [{\citenamefont {Ando}\ \emph {et~al.}(2014)\citenamefont {Ando}, \citenamefont {Hamasaki}, \citenamefont {Kurokawa}, \citenamefont {Ichiba}, \citenamefont {Yang}, \citenamefont {Novak}, \citenamefont {Sasaki}, \citenamefont {Segawa}, \citenamefont {Ando},\ and\ \citenamefont {Shiraishi}}]{Ando2014}%
  \BibitemOpen
  \bibfield  {author} {\bibinfo {author} {\bibfnamefont {Y.}~\bibnamefont {Ando}}, \bibinfo {author} {\bibfnamefont {T.}~\bibnamefont {Hamasaki}}, \bibinfo {author} {\bibfnamefont {T.}~\bibnamefont {Kurokawa}}, \bibinfo {author} {\bibfnamefont {K.}~\bibnamefont {Ichiba}}, \bibinfo {author} {\bibfnamefont {F.}~\bibnamefont {Yang}}, \bibinfo {author} {\bibfnamefont {M.}~\bibnamefont {Novak}}, \bibinfo {author} {\bibfnamefont {S.}~\bibnamefont {Sasaki}}, \bibinfo {author} {\bibfnamefont {K.}~\bibnamefont {Segawa}}, \bibinfo {author} {\bibfnamefont {Y.}~\bibnamefont {Ando}},\ and\ \bibinfo {author} {\bibfnamefont {M.}~\bibnamefont {Shiraishi}},\ }\bibfield  {title} {\bibinfo {title} {{Electrical Detection of the Spin Polarization Due to Charge Flow in the Surface State of the Topological Insulator Bi$_{1.5}$Sb$_{0.5}$Te$_{1.7}$Se$_{1.3}$}},\ }\href {https://doi.org/10.1021/nl502546c} {\bibfield  {journal} {\bibinfo  {journal} {Nano Lett.}\ }\textbf {\bibinfo {volume} {14}},\ \bibinfo {pages} {6226} (\bibinfo
  {year} {2014})}\BibitemShut {NoStop}%
\bibitem [{\citenamefont {Kondou}\ \emph {et~al.}(2016)\citenamefont {Kondou}, \citenamefont {Yoshimi}, \citenamefont {Tsukazaki}, \citenamefont {Fukuma}, \citenamefont {Matsuno}, \citenamefont {Takahashi}, \citenamefont {Kawasaki}, \citenamefont {Tokura},\ and\ \citenamefont {Otani}}]{Kondou2016}%
  \BibitemOpen
  \bibfield  {author} {\bibinfo {author} {\bibfnamefont {K.}~\bibnamefont {Kondou}}, \bibinfo {author} {\bibfnamefont {R.}~\bibnamefont {Yoshimi}}, \bibinfo {author} {\bibfnamefont {A.}~\bibnamefont {Tsukazaki}}, \bibinfo {author} {\bibfnamefont {Y.}~\bibnamefont {Fukuma}}, \bibinfo {author} {\bibfnamefont {J.}~\bibnamefont {Matsuno}}, \bibinfo {author} {\bibfnamefont {K.~S.}\ \bibnamefont {Takahashi}}, \bibinfo {author} {\bibfnamefont {M.}~\bibnamefont {Kawasaki}}, \bibinfo {author} {\bibfnamefont {Y.}~\bibnamefont {Tokura}},\ and\ \bibinfo {author} {\bibfnamefont {Y.}~\bibnamefont {Otani}},\ }\bibfield  {title} {\bibinfo {title} {{Fermi-level-dependent charge-to-spin current conversion by Dirac surface states of topological insulators}},\ }\href {https://doi.org/10.1038/nphys3833} {\bibfield  {journal} {\bibinfo  {journal} {Nat. Phys.}\ }\textbf {\bibinfo {volume} {12}},\ \bibinfo {pages} {1027} (\bibinfo {year} {2016})}\BibitemShut {NoStop}%
\bibitem [{\citenamefont {Rojas-S\'anchez}\ \emph {et~al.}(2016)\citenamefont {Rojas-S\'anchez}, \citenamefont {Oyarz\'un}, \citenamefont {Fu}, \citenamefont {Marty}, \citenamefont {Vergnaud}, \citenamefont {Gambarelli}, \citenamefont {Vila}, \citenamefont {Jamet}, \citenamefont {Ohtsubo}, \citenamefont {Taleb-Ibrahimi}, \citenamefont {Le~F\`evre}, \citenamefont {Bertran}, \citenamefont {Reyren}, \citenamefont {George},\ and\ \citenamefont {Fert}}]{Rojas2016}%
  \BibitemOpen
  \bibfield  {author} {\bibinfo {author} {\bibfnamefont {J.-C.}\ \bibnamefont {Rojas-S\'anchez}}, \bibinfo {author} {\bibfnamefont {S.}~\bibnamefont {Oyarz\'un}}, \bibinfo {author} {\bibfnamefont {Y.}~\bibnamefont {Fu}}, \bibinfo {author} {\bibfnamefont {A.}~\bibnamefont {Marty}}, \bibinfo {author} {\bibfnamefont {C.}~\bibnamefont {Vergnaud}}, \bibinfo {author} {\bibfnamefont {S.}~\bibnamefont {Gambarelli}}, \bibinfo {author} {\bibfnamefont {L.}~\bibnamefont {Vila}}, \bibinfo {author} {\bibfnamefont {M.}~\bibnamefont {Jamet}}, \bibinfo {author} {\bibfnamefont {Y.}~\bibnamefont {Ohtsubo}}, \bibinfo {author} {\bibfnamefont {A.}~\bibnamefont {Taleb-Ibrahimi}}, \bibinfo {author} {\bibfnamefont {P.}~\bibnamefont {Le~F\`evre}}, \bibinfo {author} {\bibfnamefont {F.}~\bibnamefont {Bertran}}, \bibinfo {author} {\bibfnamefont {N.}~\bibnamefont {Reyren}}, \bibinfo {author} {\bibfnamefont {J.-M.}\ \bibnamefont {George}},\ and\ \bibinfo {author} {\bibfnamefont {A.}~\bibnamefont {Fert}},\ }\bibfield  {title} {\bibinfo
  {title} {{Spin to Charge Conversion at Room Temperature by Spin Pumping into a New Type of Topological Insulator: $\ensuremath{\alpha}$-Sn Films}},\ }\href {https://doi.org/10.1103/PhysRevLett.116.096602} {\bibfield  {journal} {\bibinfo  {journal} {Phys. Rev. Lett.}\ }\textbf {\bibinfo {volume} {116}},\ \bibinfo {pages} {096602} (\bibinfo {year} {2016})}\BibitemShut {NoStop}%
\bibitem [{\citenamefont {Johansson}\ \emph {et~al.}(2018)\citenamefont {Johansson}, \citenamefont {Henk},\ and\ \citenamefont {Mertig}}]{johansson2018edelstein}%
  \BibitemOpen
  \bibfield  {author} {\bibinfo {author} {\bibfnamefont {A.}~\bibnamefont {Johansson}}, \bibinfo {author} {\bibfnamefont {J.}~\bibnamefont {Henk}},\ and\ \bibinfo {author} {\bibfnamefont {I.}~\bibnamefont {Mertig}},\ }\bibfield  {title} {\bibinfo {title} {{Edelstein effect in Weyl semimetals}},\ }\href {https://doi.org/10.1103/PhysRevB.97.085417} {\bibfield  {journal} {\bibinfo  {journal} {Phys. Rev. B}\ }\textbf {\bibinfo {volume} {97}},\ \bibinfo {pages} {085417} (\bibinfo {year} {2018})}\BibitemShut {NoStop}%
\bibitem [{\citenamefont {Zhao}\ \emph {et~al.}(2020)\citenamefont {Zhao}, \citenamefont {Karpiak}, \citenamefont {Khokhriakov}, \citenamefont {Johansson}, \citenamefont {Hoque}, \citenamefont {Xu}, \citenamefont {Jiang}, \citenamefont {Mertig},\ and\ \citenamefont {Dash}}]{Zhao2020}%
  \BibitemOpen
  \bibfield  {author} {\bibinfo {author} {\bibfnamefont {B.}~\bibnamefont {Zhao}}, \bibinfo {author} {\bibfnamefont {B.}~\bibnamefont {Karpiak}}, \bibinfo {author} {\bibfnamefont {D.}~\bibnamefont {Khokhriakov}}, \bibinfo {author} {\bibfnamefont {A.}~\bibnamefont {Johansson}}, \bibinfo {author} {\bibfnamefont {A.~M.}\ \bibnamefont {Hoque}}, \bibinfo {author} {\bibfnamefont {X.}~\bibnamefont {Xu}}, \bibinfo {author} {\bibfnamefont {Y.}~\bibnamefont {Jiang}}, \bibinfo {author} {\bibfnamefont {I.}~\bibnamefont {Mertig}},\ and\ \bibinfo {author} {\bibfnamefont {S.~P.}\ \bibnamefont {Dash}},\ }\bibfield  {title} {\bibinfo {title} {{Unconventional Charge–Spin Conversion in Weyl‐Semimetal WTe$_2$}},\ }\bibfield  {journal} {\bibinfo  {journal} {Advanced Materials}\ }\textbf {\bibinfo {volume} {32}},\ \href {https://doi.org/10.1002/adma.202000818} {10.1002/adma.202000818} (\bibinfo {year} {2020})\BibitemShut {NoStop}%
\bibitem [{\citenamefont {Shalygin}\ \emph {et~al.}(2012)\citenamefont {Shalygin}, \citenamefont {Sofronov}, \citenamefont {Vorob'ev},\ and\ \citenamefont {Farbshtein}}]{Shalygin2012}%
  \BibitemOpen
  \bibfield  {author} {\bibinfo {author} {\bibfnamefont {V.~A.}\ \bibnamefont {Shalygin}}, \bibinfo {author} {\bibfnamefont {A.~N.}\ \bibnamefont {Sofronov}}, \bibinfo {author} {\bibfnamefont {L.~E.}\ \bibnamefont {Vorob'ev}},\ and\ \bibinfo {author} {\bibfnamefont {I.~I.}\ \bibnamefont {Farbshtein}},\ }\bibfield  {title} {\bibinfo {title} {{Current-induced spin polarization of holes in tellurium}},\ }\href {https://doi.org/10.1134/s1063783412120281} {\bibfield  {journal} {\bibinfo  {journal} {Physics of the Solid State}\ }\textbf {\bibinfo {volume} {54}},\ \bibinfo {pages} {2362} (\bibinfo {year} {2012})}\BibitemShut {NoStop}%
\bibitem [{\citenamefont {Furukawa}\ \emph {et~al.}(2017)\citenamefont {Furukawa}, \citenamefont {Shimokawa}, \citenamefont {Kobayashi},\ and\ \citenamefont {Itou}}]{Furukawa2017}%
  \BibitemOpen
  \bibfield  {author} {\bibinfo {author} {\bibfnamefont {T.}~\bibnamefont {Furukawa}}, \bibinfo {author} {\bibfnamefont {Y.}~\bibnamefont {Shimokawa}}, \bibinfo {author} {\bibfnamefont {K.}~\bibnamefont {Kobayashi}},\ and\ \bibinfo {author} {\bibfnamefont {T.}~\bibnamefont {Itou}},\ }\bibfield  {title} {\bibinfo {title} {{Observation of current-induced bulk magnetization in elemental tellurium}},\ }\href {https://doi.org/10.1038/s41467-017-01093-3} {\bibfield  {journal} {\bibinfo  {journal} {Nature Communications}\ }\textbf {\bibinfo {volume} {8}},\ \bibinfo {pages} {954} (\bibinfo {year} {2017})}\BibitemShut {NoStop}%
\bibitem [{\citenamefont {Furukawa}\ \emph {et~al.}(2021)\citenamefont {Furukawa}, \citenamefont {Watanabe}, \citenamefont {Ogasawara}, \citenamefont {Kobayashi},\ and\ \citenamefont {Itou}}]{Furukawa2021}%
  \BibitemOpen
  \bibfield  {author} {\bibinfo {author} {\bibfnamefont {T.}~\bibnamefont {Furukawa}}, \bibinfo {author} {\bibfnamefont {Y.}~\bibnamefont {Watanabe}}, \bibinfo {author} {\bibfnamefont {N.}~\bibnamefont {Ogasawara}}, \bibinfo {author} {\bibfnamefont {K.}~\bibnamefont {Kobayashi}},\ and\ \bibinfo {author} {\bibfnamefont {T.}~\bibnamefont {Itou}},\ }\bibfield  {title} {\bibinfo {title} {{Current-induced magnetization caused by crystal chirality in nonmagnetic elemental tellurium}},\ }\href {https://doi.org/10.1103/PhysRevResearch.3.023111} {\bibfield  {journal} {\bibinfo  {journal} {Phys. Rev. Res.}\ }\textbf {\bibinfo {volume} {3}},\ \bibinfo {pages} {023111} (\bibinfo {year} {2021})}\BibitemShut {NoStop}%
\bibitem [{\citenamefont {Yang}\ \emph {et~al.}(2021{\natexlab{a}})\citenamefont {Yang}, \citenamefont {Naaman}, \citenamefont {Paltiel},\ and\ \citenamefont {Parkin}}]{yang2021chiral}%
  \BibitemOpen
  \bibfield  {author} {\bibinfo {author} {\bibfnamefont {S.-H.}\ \bibnamefont {Yang}}, \bibinfo {author} {\bibfnamefont {R.}~\bibnamefont {Naaman}}, \bibinfo {author} {\bibfnamefont {Y.}~\bibnamefont {Paltiel}},\ and\ \bibinfo {author} {\bibfnamefont {S.~S.}\ \bibnamefont {Parkin}},\ }\bibfield  {title} {\bibinfo {title} {{Chiral spintronics}},\ }\href {https://doi.org/10.1038/s42254-021-00302-9} {\bibfield  {journal} {\bibinfo  {journal} {Nature Reviews Physics}\ }\textbf {\bibinfo {volume} {3}},\ \bibinfo {pages} {328} (\bibinfo {year} {2021}{\natexlab{a}})}\BibitemShut {NoStop}%
\bibitem [{\citenamefont {Calavalle}\ \emph {et~al.}(2022)\citenamefont {Calavalle}, \citenamefont {Suárez-Rodríguez}, \citenamefont {Martín-García}, \citenamefont {Johansson}, \citenamefont {Vaz}, \citenamefont {Yang}, \citenamefont {Maznichenko}, \citenamefont {Ostanin}, \citenamefont {Mateo-Alonso}, \citenamefont {Chuvilin}, \citenamefont {Mertig}, \citenamefont {Gobbi}, \citenamefont {Casanova},\ and\ \citenamefont {Hueso}}]{Calavalle2022}%
  \BibitemOpen
  \bibfield  {author} {\bibinfo {author} {\bibfnamefont {F.}~\bibnamefont {Calavalle}}, \bibinfo {author} {\bibfnamefont {M.}~\bibnamefont {Suárez-Rodríguez}}, \bibinfo {author} {\bibfnamefont {B.}~\bibnamefont {Martín-García}}, \bibinfo {author} {\bibfnamefont {A.}~\bibnamefont {Johansson}}, \bibinfo {author} {\bibfnamefont {D.~C.}\ \bibnamefont {Vaz}}, \bibinfo {author} {\bibfnamefont {H.}~\bibnamefont {Yang}}, \bibinfo {author} {\bibfnamefont {I.~V.}\ \bibnamefont {Maznichenko}}, \bibinfo {author} {\bibfnamefont {S.}~\bibnamefont {Ostanin}}, \bibinfo {author} {\bibfnamefont {A.}~\bibnamefont {Mateo-Alonso}}, \bibinfo {author} {\bibfnamefont {A.}~\bibnamefont {Chuvilin}}, \bibinfo {author} {\bibfnamefont {I.}~\bibnamefont {Mertig}}, \bibinfo {author} {\bibfnamefont {M.}~\bibnamefont {Gobbi}}, \bibinfo {author} {\bibfnamefont {F.}~\bibnamefont {Casanova}},\ and\ \bibinfo {author} {\bibfnamefont {L.~E.}\ \bibnamefont {Hueso}},\ }\bibfield  {title} {\bibinfo {title} {{Gate-tuneable and chirality-dependent
  charge-to-spin conversion in tellurium nanowires}},\ }\href {https://doi.org/0.1038/s41563-022-01211-7} {\bibfield  {journal} {\bibinfo  {journal} {Nat. Mater.}\ }\textbf {\bibinfo {volume} {21}},\ \bibinfo {pages} {526} (\bibinfo {year} {2022})}\BibitemShut {NoStop}%
\bibitem [{\citenamefont {G{\"o}bel}\ \emph {et~al.}(2025)\citenamefont {G{\"o}bel}, \citenamefont {Schimpf},\ and\ \citenamefont {Mertig}}]{gobel2025chirality}%
  \BibitemOpen
  \bibfield  {author} {\bibinfo {author} {\bibfnamefont {B.}~\bibnamefont {G{\"o}bel}}, \bibinfo {author} {\bibfnamefont {L.}~\bibnamefont {Schimpf}},\ and\ \bibinfo {author} {\bibfnamefont {I.}~\bibnamefont {Mertig}},\ }\bibfield  {title} {\bibinfo {title} {{Chirality-induced orbital Edelstein effect in an analytically solvable model}},\ }\href {https://doi.org/10.48550/arXiv.2502.04978} {\bibfield  {journal} {\bibinfo  {journal} {arXiv preprint arXiv:2502.04978}\ } (\bibinfo {year} {2025})}\BibitemShut {NoStop}%
\bibitem [{\citenamefont {Vaz}\ \emph {et~al.}(2019)\citenamefont {Vaz}, \citenamefont {No\"{e}l}, \citenamefont {Johansson}, \citenamefont {G\"{o}bel}, \citenamefont {Bruno}, \citenamefont {Singh}, \citenamefont {McKeown-Walker}, \citenamefont {Trier}, \citenamefont {Vicente-Arche}, \citenamefont {Sander}, \citenamefont {Valencia}, \citenamefont {Bruneel}, \citenamefont {Vivek}, \citenamefont {Gabay}, \citenamefont {Bergeal}, \citenamefont {Baumberger}, \citenamefont {Okuno}, \citenamefont {Barth\'{e}l\'{e}my}, \citenamefont {Fert}, \citenamefont {Vila}, \citenamefont {Mertig}, \citenamefont {Attan\'{e}},\ and\ \citenamefont {Bibes}}]{Vaz2019}%
  \BibitemOpen
  \bibfield  {author} {\bibinfo {author} {\bibfnamefont {D.~C.}\ \bibnamefont {Vaz}}, \bibinfo {author} {\bibfnamefont {P.}~\bibnamefont {No\"{e}l}}, \bibinfo {author} {\bibfnamefont {A.}~\bibnamefont {Johansson}}, \bibinfo {author} {\bibfnamefont {B.}~\bibnamefont {G\"{o}bel}}, \bibinfo {author} {\bibfnamefont {F.~Y.}\ \bibnamefont {Bruno}}, \bibinfo {author} {\bibfnamefont {G.}~\bibnamefont {Singh}}, \bibinfo {author} {\bibfnamefont {S.}~\bibnamefont {McKeown-Walker}}, \bibinfo {author} {\bibfnamefont {F.}~\bibnamefont {Trier}}, \bibinfo {author} {\bibfnamefont {L.~M.}\ \bibnamefont {Vicente-Arche}}, \bibinfo {author} {\bibfnamefont {A.}~\bibnamefont {Sander}}, \bibinfo {author} {\bibfnamefont {S.}~\bibnamefont {Valencia}}, \bibinfo {author} {\bibfnamefont {P.}~\bibnamefont {Bruneel}}, \bibinfo {author} {\bibfnamefont {M.}~\bibnamefont {Vivek}}, \bibinfo {author} {\bibfnamefont {M.}~\bibnamefont {Gabay}}, \bibinfo {author} {\bibfnamefont {N.}~\bibnamefont {Bergeal}}, \bibinfo {author} {\bibfnamefont
  {F.}~\bibnamefont {Baumberger}}, \bibinfo {author} {\bibfnamefont {H.}~\bibnamefont {Okuno}}, \bibinfo {author} {\bibfnamefont {A.}~\bibnamefont {Barth\'{e}l\'{e}my}}, \bibinfo {author} {\bibfnamefont {A.}~\bibnamefont {Fert}}, \bibinfo {author} {\bibfnamefont {L.}~\bibnamefont {Vila}}, \bibinfo {author} {\bibfnamefont {I.}~\bibnamefont {Mertig}}, \bibinfo {author} {\bibfnamefont {J.-P.}\ \bibnamefont {Attan\'{e}}},\ and\ \bibinfo {author} {\bibfnamefont {M.}~\bibnamefont {Bibes}},\ }\bibfield  {title} {\bibinfo {title} {{Mapping spin–charge conversion to the band structure in a topological oxide two-dimensional electron gas}},\ }\href {https://doi.org/10.1038/s41563-019-0467-4} {\bibfield  {journal} {\bibinfo  {journal} {Nature Materials}\ }\textbf {\bibinfo {volume} {18}},\ \bibinfo {pages} {1187} (\bibinfo {year} {2019})}\BibitemShut {NoStop}%
\bibitem [{\citenamefont {Johansson}\ \emph {et~al.}(2021)\citenamefont {Johansson}, \citenamefont {G\"obel}, \citenamefont {Henk}, \citenamefont {Bibes},\ and\ \citenamefont {Mertig}}]{Johansson2021}%
  \BibitemOpen
  \bibfield  {author} {\bibinfo {author} {\bibfnamefont {A.}~\bibnamefont {Johansson}}, \bibinfo {author} {\bibfnamefont {B.}~\bibnamefont {G\"obel}}, \bibinfo {author} {\bibfnamefont {J.}~\bibnamefont {Henk}}, \bibinfo {author} {\bibfnamefont {M.}~\bibnamefont {Bibes}},\ and\ \bibinfo {author} {\bibfnamefont {I.}~\bibnamefont {Mertig}},\ }\bibfield  {title} {\bibinfo {title} {{Spin and orbital Edelstein effects in a two-dimensional electron gas: Theory and application to SrTiO$_3$ interfaces}},\ }\href {https://doi.org/10.1103/PhysRevResearch.3.013275} {\bibfield  {journal} {\bibinfo  {journal} {Phys. Rev. Res.}\ }\textbf {\bibinfo {volume} {3}},\ \bibinfo {pages} {013275} (\bibinfo {year} {2021})}\BibitemShut {NoStop}%
\bibitem [{\citenamefont {Trama}\ \emph {et~al.}(2022)\citenamefont {Trama}, \citenamefont {Cataudella}, \citenamefont {Perroni}, \citenamefont {Romeo},\ and\ \citenamefont {Citro}}]{trama2022tunable}%
  \BibitemOpen
  \bibfield  {author} {\bibinfo {author} {\bibfnamefont {M.}~\bibnamefont {Trama}}, \bibinfo {author} {\bibfnamefont {V.}~\bibnamefont {Cataudella}}, \bibinfo {author} {\bibfnamefont {C.~A.}\ \bibnamefont {Perroni}}, \bibinfo {author} {\bibfnamefont {F.}~\bibnamefont {Romeo}},\ and\ \bibinfo {author} {\bibfnamefont {R.}~\bibnamefont {Citro}},\ }\bibfield  {title} {\bibinfo {title} {{Tunable spin and orbital Edelstein effect at (111) LaAlO$_3$/SrTiO$_3$ interface}},\ }\href {https://doi.org/10.3390/nano12142494} {\bibfield  {journal} {\bibinfo  {journal} {Nanomaterials}\ }\textbf {\bibinfo {volume} {12}},\ \bibinfo {pages} {2494} (\bibinfo {year} {2022})}\BibitemShut {NoStop}%
\bibitem [{\citenamefont {Varotto}\ \emph {et~al.}(2022)\citenamefont {Varotto}, \citenamefont {Johansson}, \citenamefont {Göbel}, \citenamefont {Vicente-Arche}, \citenamefont {Mallik}, \citenamefont {Bréhin}, \citenamefont {Salazar}, \citenamefont {Bertran}, \citenamefont {Fèvre}, \citenamefont {Bergeal}, \citenamefont {Rault}, \citenamefont {Mertig},\ and\ \citenamefont {Bibes}}]{varotto2022direct}%
  \BibitemOpen
  \bibfield  {author} {\bibinfo {author} {\bibfnamefont {S.}~\bibnamefont {Varotto}}, \bibinfo {author} {\bibfnamefont {A.}~\bibnamefont {Johansson}}, \bibinfo {author} {\bibfnamefont {B.}~\bibnamefont {Göbel}}, \bibinfo {author} {\bibfnamefont {L.~M.}\ \bibnamefont {Vicente-Arche}}, \bibinfo {author} {\bibfnamefont {S.}~\bibnamefont {Mallik}}, \bibinfo {author} {\bibfnamefont {J.}~\bibnamefont {Bréhin}}, \bibinfo {author} {\bibfnamefont {R.}~\bibnamefont {Salazar}}, \bibinfo {author} {\bibfnamefont {F.}~\bibnamefont {Bertran}}, \bibinfo {author} {\bibfnamefont {P.~L.}\ \bibnamefont {Fèvre}}, \bibinfo {author} {\bibfnamefont {N.}~\bibnamefont {Bergeal}}, \bibinfo {author} {\bibfnamefont {J.}~\bibnamefont {Rault}}, \bibinfo {author} {\bibfnamefont {I.}~\bibnamefont {Mertig}},\ and\ \bibinfo {author} {\bibfnamefont {M.}~\bibnamefont {Bibes}},\ }\bibfield  {title} {\bibinfo {title} {{Direct visualization of Rashba-split bands and spin/orbital-charge interconversion at KTaO$_3$ interfaces}},\ }\href
  {https://doi.org/10.1038/s41467-022-33621-1} {\bibfield  {journal} {\bibinfo  {journal} {Nat. Commun.}\ }\textbf {\bibinfo {volume} {13}},\ \bibinfo {pages} {6165} (\bibinfo {year} {2022})}\BibitemShut {NoStop}%
\bibitem [{\citenamefont {Krishnia}\ \emph {et~al.}(2024)\citenamefont {Krishnia}, \citenamefont {Voj{\'a}{\v{c}}ek}, \citenamefont {Gomes}, \citenamefont {Sebe}, \citenamefont {Ibrahim}, \citenamefont {Li}, \citenamefont {Vicente-Arche}, \citenamefont {Collin}, \citenamefont {Denneulin}, \citenamefont {Dunin-Borkowski} \emph {et~al.}}]{krishnia2024interfacial}%
  \BibitemOpen
  \bibfield  {author} {\bibinfo {author} {\bibfnamefont {S.}~\bibnamefont {Krishnia}}, \bibinfo {author} {\bibfnamefont {L.}~\bibnamefont {Voj{\'a}{\v{c}}ek}}, \bibinfo {author} {\bibfnamefont {T.~D. C. S.~C.}\ \bibnamefont {Gomes}}, \bibinfo {author} {\bibfnamefont {N.}~\bibnamefont {Sebe}}, \bibinfo {author} {\bibfnamefont {F.}~\bibnamefont {Ibrahim}}, \bibinfo {author} {\bibfnamefont {J.}~\bibnamefont {Li}}, \bibinfo {author} {\bibfnamefont {L.~M.}\ \bibnamefont {Vicente-Arche}}, \bibinfo {author} {\bibfnamefont {S.}~\bibnamefont {Collin}}, \bibinfo {author} {\bibfnamefont {T.}~\bibnamefont {Denneulin}}, \bibinfo {author} {\bibfnamefont {R.~E.}\ \bibnamefont {Dunin-Borkowski}}, \emph {et~al.},\ }\bibfield  {title} {\bibinfo {title} {{Interfacial spin-orbitronic effects controlled with different oxidation levels at the {Co{\textbar}Al} interface}},\ }\bibfield  {journal} {\bibinfo  {journal} {arXiv preprint arXiv:2409.10685}\ }\href {https://doi.org/10.48550/arXiv.2409.10685} {10.48550/arXiv.2409.10685}
  (\bibinfo {year} {2024})\BibitemShut {NoStop}%
\bibitem [{\citenamefont {Katsantonis}\ \emph {et~al.}(2023)\citenamefont {Katsantonis}, \citenamefont {Tasolamprou}, \citenamefont {Koschny}, \citenamefont {Economou}, \citenamefont {Kafesaki},\ and\ \citenamefont {Valagiannopoulos}}]{katsantonis2023giant}%
  \BibitemOpen
  \bibfield  {author} {\bibinfo {author} {\bibfnamefont {I.}~\bibnamefont {Katsantonis}}, \bibinfo {author} {\bibfnamefont {A.~C.}\ \bibnamefont {Tasolamprou}}, \bibinfo {author} {\bibfnamefont {T.}~\bibnamefont {Koschny}}, \bibinfo {author} {\bibfnamefont {E.~N.}\ \bibnamefont {Economou}}, \bibinfo {author} {\bibfnamefont {M.}~\bibnamefont {Kafesaki}},\ and\ \bibinfo {author} {\bibfnamefont {C.}~\bibnamefont {Valagiannopoulos}},\ }\bibfield  {title} {\bibinfo {title} {Giant enhancement of nonreciprocity in gyrotropic heterostructures},\ }\href {https://doi.org/10.1038/s41598-023-48503-9} {\bibfield  {journal} {\bibinfo  {journal} {Scientific Reports}\ }\textbf {\bibinfo {volume} {13}},\ \bibinfo {pages} {21986} (\bibinfo {year} {2023})}\BibitemShut {NoStop}%
\bibitem [{\citenamefont {Orazbay}\ and\ \citenamefont {Valagiannopoulos}(2024)}]{orazbay2024twistronics}%
  \BibitemOpen
  \bibfield  {author} {\bibinfo {author} {\bibfnamefont {M.}~\bibnamefont {Orazbay}}\ and\ \bibinfo {author} {\bibfnamefont {C.}~\bibnamefont {Valagiannopoulos}},\ }\bibfield  {title} {\bibinfo {title} {Twistronics-based polarization engineering},\ }\href {https://doi.org/10.1103/PhysRevApplied.22.054028} {\bibfield  {journal} {\bibinfo  {journal} {Phys. Rev. Appl.}\ }\textbf {\bibinfo {volume} {22}},\ \bibinfo {pages} {054028} (\bibinfo {year} {2024})}\BibitemShut {NoStop}%
\bibitem [{\citenamefont {Pari}\ \emph {et~al.}(2025)\citenamefont {Pari}, \citenamefont {Jaeschke-Ubiergo}, \citenamefont {Chakraborty}, \citenamefont {\ifmmode~\check{S}\else \v{S}\fi{}mejkal},\ and\ \citenamefont {Sinova}}]{pari2025nonrelativistic}%
  \BibitemOpen
  \bibfield  {author} {\bibinfo {author} {\bibfnamefont {N.~A.~A.}\ \bibnamefont {Pari}}, \bibinfo {author} {\bibfnamefont {R.}~\bibnamefont {Jaeschke-Ubiergo}}, \bibinfo {author} {\bibfnamefont {A.}~\bibnamefont {Chakraborty}}, \bibinfo {author} {\bibfnamefont {L.}~\bibnamefont {\ifmmode~\check{S}\else \v{S}\fi{}mejkal}},\ and\ \bibinfo {author} {\bibfnamefont {J.}~\bibnamefont {Sinova}},\ }\bibfield  {title} {\bibinfo {title} {{Nonrelativistic linear Edelstein effect in helical ${\mathrm{EuIn}}_{2}{\mathrm{As}}_{2}$}},\ }\href {https://doi.org/10.1103/k9p4-tfhd} {\bibfield  {journal} {\bibinfo  {journal} {Phys. Rev. B}\ }\textbf {\bibinfo {volume} {112}},\ \bibinfo {pages} {024404} (\bibinfo {year} {2025})}\BibitemShut {NoStop}%
\bibitem [{\citenamefont {Chakraborty}\ \emph {et~al.}(2025)\citenamefont {Chakraborty}, \citenamefont {Hellenes}, \citenamefont {Jaeschke-Ubiergo}, \citenamefont {Jungwirth}, \citenamefont {{\v{S}}mejkal},\ and\ \citenamefont {Sinova}}]{chakraborty2024highly}%
  \BibitemOpen
  \bibfield  {author} {\bibinfo {author} {\bibfnamefont {A.}~\bibnamefont {Chakraborty}}, \bibinfo {author} {\bibfnamefont {A.~B.}\ \bibnamefont {Hellenes}}, \bibinfo {author} {\bibfnamefont {R.}~\bibnamefont {Jaeschke-Ubiergo}}, \bibinfo {author} {\bibfnamefont {T.}~\bibnamefont {Jungwirth}}, \bibinfo {author} {\bibfnamefont {L.}~\bibnamefont {{\v{S}}mejkal}},\ and\ \bibinfo {author} {\bibfnamefont {J.}~\bibnamefont {Sinova}},\ }\bibfield  {title} {\bibinfo {title} {{Highly efficient non-relativistic Edelstein effect in nodal p-wave magnets}},\ }\href {https://doi.org/10.1038/s41467-025-62516-0} {\bibfield  {journal} {\bibinfo  {journal} {Nature Communications}\ }\textbf {\bibinfo {volume} {16}},\ \bibinfo {pages} {7270} (\bibinfo {year} {2025})}\BibitemShut {NoStop}%
\bibitem [{\citenamefont {Cysne}\ \emph {et~al.}(2021)\citenamefont {Cysne}, \citenamefont {Guimar\~aes}, \citenamefont {Canonico}, \citenamefont {Rappoport},\ and\ \citenamefont {Muniz}}]{Cysne2021nano}%
  \BibitemOpen
  \bibfield  {author} {\bibinfo {author} {\bibfnamefont {T.~P.}\ \bibnamefont {Cysne}}, \bibinfo {author} {\bibfnamefont {F.~S.~M.}\ \bibnamefont {Guimar\~aes}}, \bibinfo {author} {\bibfnamefont {L.~M.}\ \bibnamefont {Canonico}}, \bibinfo {author} {\bibfnamefont {T.~G.}\ \bibnamefont {Rappoport}},\ and\ \bibinfo {author} {\bibfnamefont {R.~B.}\ \bibnamefont {Muniz}},\ }\bibfield  {title} {\bibinfo {title} {{Orbital magnetoelectric effect in zigzag nanoribbons of $p$-band systems}},\ }\href {https://doi.org/10.1103/PhysRevB.104.165403} {\bibfield  {journal} {\bibinfo  {journal} {Phys. Rev. B}\ }\textbf {\bibinfo {volume} {104}},\ \bibinfo {pages} {165403} (\bibinfo {year} {2021})}\BibitemShut {NoStop}%
\bibitem [{\citenamefont {Lee}\ \emph {et~al.}(2022)\citenamefont {Lee}, \citenamefont {de~Sousa}, \citenamefont {Kwon}, \citenamefont {de~Juan}, \citenamefont {Chi}, \citenamefont {Casanova},\ and\ \citenamefont {Low}}]{Lee2022}%
  \BibitemOpen
  \bibfield  {author} {\bibinfo {author} {\bibfnamefont {S.}~\bibnamefont {Lee}}, \bibinfo {author} {\bibfnamefont {D.~J.~P.}\ \bibnamefont {de~Sousa}}, \bibinfo {author} {\bibfnamefont {Y.-K.}\ \bibnamefont {Kwon}}, \bibinfo {author} {\bibfnamefont {F.}~\bibnamefont {de~Juan}}, \bibinfo {author} {\bibfnamefont {Z.}~\bibnamefont {Chi}}, \bibinfo {author} {\bibfnamefont {F.}~\bibnamefont {Casanova}},\ and\ \bibinfo {author} {\bibfnamefont {T.}~\bibnamefont {Low}},\ }\bibfield  {title} {\bibinfo {title} {Charge-to-spin conversion in twisted $\mathrm{graphene}/{\mathrm{wse}}_{2}$ heterostructures},\ }\href {https://doi.org/10.1103/PhysRevB.106.165420} {\bibfield  {journal} {\bibinfo  {journal} {Phys. Rev. B}\ }\textbf {\bibinfo {volume} {106}},\ \bibinfo {pages} {165420} (\bibinfo {year} {2022})}\BibitemShut {NoStop}%
\bibitem [{\citenamefont {Ingla-Ayn{\'{e}}s}\ \emph {et~al.}(2022)\citenamefont {Ingla-Ayn{\'{e}}s}, \citenamefont {Groen}, \citenamefont {Herling}, \citenamefont {Ontoso}, \citenamefont {Safeer}, \citenamefont {de~Juan}, \citenamefont {Hueso}, \citenamefont {Gobbi},\ and\ \citenamefont {Casanova}}]{InglaAyns2022}%
  \BibitemOpen
  \bibfield  {author} {\bibinfo {author} {\bibfnamefont {J.}~\bibnamefont {Ingla-Ayn{\'{e}}s}}, \bibinfo {author} {\bibfnamefont {I.}~\bibnamefont {Groen}}, \bibinfo {author} {\bibfnamefont {F.}~\bibnamefont {Herling}}, \bibinfo {author} {\bibfnamefont {N.}~\bibnamefont {Ontoso}}, \bibinfo {author} {\bibfnamefont {C.~K.}\ \bibnamefont {Safeer}}, \bibinfo {author} {\bibfnamefont {F.}~\bibnamefont {de~Juan}}, \bibinfo {author} {\bibfnamefont {L.~E.}\ \bibnamefont {Hueso}}, \bibinfo {author} {\bibfnamefont {M.}~\bibnamefont {Gobbi}},\ and\ \bibinfo {author} {\bibfnamefont {F.}~\bibnamefont {Casanova}},\ }\bibfield  {title} {\bibinfo {title} {{Omnidirectional spin-to-charge conversion in graphene/NbSe$_2$ van der Waals heterostructures}},\ }\href {https://doi.org/10.1088/2053-1583/ac76d1} {\bibfield  {journal} {\bibinfo  {journal} {2D Materials}\ }\textbf {\bibinfo {volume} {9}},\ \bibinfo {pages} {045001} (\bibinfo {year} {2022})}\BibitemShut {NoStop}%
\bibitem [{\citenamefont {Cysne}\ \emph {et~al.}(2023)\citenamefont {Cysne}, \citenamefont {Guimar\~aes}, \citenamefont {Canonico}, \citenamefont {Costa}, \citenamefont {Rappoport},\ and\ \citenamefont {Muniz}}]{Cysne2023}%
  \BibitemOpen
  \bibfield  {author} {\bibinfo {author} {\bibfnamefont {T.~P.}\ \bibnamefont {Cysne}}, \bibinfo {author} {\bibfnamefont {F.~S.~M.}\ \bibnamefont {Guimar\~aes}}, \bibinfo {author} {\bibfnamefont {L.~M.}\ \bibnamefont {Canonico}}, \bibinfo {author} {\bibfnamefont {M.}~\bibnamefont {Costa}}, \bibinfo {author} {\bibfnamefont {T.~G.}\ \bibnamefont {Rappoport}},\ and\ \bibinfo {author} {\bibfnamefont {R.~B.}\ \bibnamefont {Muniz}},\ }\bibfield  {title} {\bibinfo {title} {{Orbital magnetoelectric effect in nanoribbons of transition metal dichalcogenides}},\ }\href {https://doi.org/10.1103/PhysRevB.107.115402} {\bibfield  {journal} {\bibinfo  {journal} {Phys. Rev. B}\ }\textbf {\bibinfo {volume} {107}},\ \bibinfo {pages} {115402} (\bibinfo {year} {2023})}\BibitemShut {NoStop}%
\bibitem [{\citenamefont {Chirolli}\ \emph {et~al.}(2022)\citenamefont {Chirolli}, \citenamefont {Mercaldo}, \citenamefont {Guarcello}, \citenamefont {Giazotto},\ and\ \citenamefont {Cuoco}}]{Chirolli2022}%
  \BibitemOpen
  \bibfield  {author} {\bibinfo {author} {\bibfnamefont {L.}~\bibnamefont {Chirolli}}, \bibinfo {author} {\bibfnamefont {M.~T.}\ \bibnamefont {Mercaldo}}, \bibinfo {author} {\bibfnamefont {C.}~\bibnamefont {Guarcello}}, \bibinfo {author} {\bibfnamefont {F.}~\bibnamefont {Giazotto}},\ and\ \bibinfo {author} {\bibfnamefont {M.}~\bibnamefont {Cuoco}},\ }\bibfield  {title} {\bibinfo {title} {{Colossal Orbital Edelstein Effect in Noncentrosymmetric Superconductors}},\ }\href {https://doi.org/10.1103/PhysRevLett.128.217703} {\bibfield  {journal} {\bibinfo  {journal} {Phys. Rev. Lett.}\ }\textbf {\bibinfo {volume} {128}},\ \bibinfo {pages} {217703} (\bibinfo {year} {2022})}\BibitemShut {NoStop}%
\bibitem [{\citenamefont {Han}\ \emph {et~al.}(2018)\citenamefont {Han}, \citenamefont {Otani},\ and\ \citenamefont {Maekawa}}]{Han2018}%
  \BibitemOpen
  \bibfield  {author} {\bibinfo {author} {\bibfnamefont {W.}~\bibnamefont {Han}}, \bibinfo {author} {\bibfnamefont {Y.}~\bibnamefont {Otani}},\ and\ \bibinfo {author} {\bibfnamefont {S.}~\bibnamefont {Maekawa}},\ }\bibfield  {title} {\bibinfo {title} {{Quantum materials for spin and charge conversion}},\ }\href {https://doi.org/10.1038/s41535-018-0100-9} {\bibfield  {journal} {\bibinfo  {journal} {npj Quantum Materials}\ }\textbf {\bibinfo {volume} {3}},\ \bibinfo {pages} {27} (\bibinfo {year} {2018})}\BibitemShut {NoStop}%
\bibitem [{\citenamefont {Thonhauser}(2011)}]{thonhauser2011theory}%
  \BibitemOpen
  \bibfield  {author} {\bibinfo {author} {\bibfnamefont {T.}~\bibnamefont {Thonhauser}},\ }\bibfield  {title} {\bibinfo {title} {{Theory of orbital magnetization in solids}},\ }\href {https://doi.org/10.1142/S0217979211058912} {\bibfield  {journal} {\bibinfo  {journal} {International Journal of Modern Physics B}\ }\textbf {\bibinfo {volume} {25}},\ \bibinfo {pages} {1429} (\bibinfo {year} {2011})}\BibitemShut {NoStop}%
\bibitem [{\citenamefont {Park}\ \emph {et~al.}(2011)\citenamefont {Park}, \citenamefont {Kim}, \citenamefont {Yu}, \citenamefont {Han},\ and\ \citenamefont {Kim}}]{Park2011}%
  \BibitemOpen
  \bibfield  {author} {\bibinfo {author} {\bibfnamefont {S.~R.}\ \bibnamefont {Park}}, \bibinfo {author} {\bibfnamefont {C.~H.}\ \bibnamefont {Kim}}, \bibinfo {author} {\bibfnamefont {J.}~\bibnamefont {Yu}}, \bibinfo {author} {\bibfnamefont {J.~H.}\ \bibnamefont {Han}},\ and\ \bibinfo {author} {\bibfnamefont {C.}~\bibnamefont {Kim}},\ }\bibfield  {title} {\bibinfo {title} {{Orbital-Angular-Momentum Based Origin of Rashba-Type Surface Band Splitting}},\ }\href {https://doi.org/10.1103/PhysRevLett.107.156803} {\bibfield  {journal} {\bibinfo  {journal} {Phys. Rev. Lett.}\ }\textbf {\bibinfo {volume} {107}},\ \bibinfo {pages} {156803} (\bibinfo {year} {2011})}\BibitemShut {NoStop}%
\bibitem [{\citenamefont {Park}\ \emph {et~al.}(2012)\citenamefont {Park}, \citenamefont {Kim}, \citenamefont {Rhim},\ and\ \citenamefont {Han}}]{Park2012}%
  \BibitemOpen
  \bibfield  {author} {\bibinfo {author} {\bibfnamefont {J.-H.}\ \bibnamefont {Park}}, \bibinfo {author} {\bibfnamefont {C.~H.}\ \bibnamefont {Kim}}, \bibinfo {author} {\bibfnamefont {J.-W.}\ \bibnamefont {Rhim}},\ and\ \bibinfo {author} {\bibfnamefont {J.~H.}\ \bibnamefont {Han}},\ }\bibfield  {title} {\bibinfo {title} {{Orbital Rashba effect and its detection by circular dichroism angle-resolved photoemission spectroscopy}},\ }\href {https://doi.org/10.1103/PhysRevB.85.195401} {\bibfield  {journal} {\bibinfo  {journal} {Phys. Rev. B}\ }\textbf {\bibinfo {volume} {85}},\ \bibinfo {pages} {195401} (\bibinfo {year} {2012})}\BibitemShut {NoStop}%
\bibitem [{\citenamefont {Go}\ \emph {et~al.}(2021)\citenamefont {Go}, \citenamefont {Jo}, \citenamefont {Gao}, \citenamefont {Ando}, \citenamefont {Bl\"ugel}, \citenamefont {Lee},\ and\ \citenamefont {Mokrousov}}]{Go2021}%
  \BibitemOpen
  \bibfield  {author} {\bibinfo {author} {\bibfnamefont {D.}~\bibnamefont {Go}}, \bibinfo {author} {\bibfnamefont {D.}~\bibnamefont {Jo}}, \bibinfo {author} {\bibfnamefont {T.}~\bibnamefont {Gao}}, \bibinfo {author} {\bibfnamefont {K.}~\bibnamefont {Ando}}, \bibinfo {author} {\bibfnamefont {S.}~\bibnamefont {Bl\"ugel}}, \bibinfo {author} {\bibfnamefont {H.-W.}\ \bibnamefont {Lee}},\ and\ \bibinfo {author} {\bibfnamefont {Y.}~\bibnamefont {Mokrousov}},\ }\bibfield  {title} {\bibinfo {title} {{Orbital Rashba effect in a surface-oxidized Cu film}},\ }\href {https://doi.org/10.1103/PhysRevB.103.L121113} {\bibfield  {journal} {\bibinfo  {journal} {Phys. Rev. B}\ }\textbf {\bibinfo {volume} {103}},\ \bibinfo {pages} {L121113} (\bibinfo {year} {2021})}\BibitemShut {NoStop}%
\bibitem [{\citenamefont {Johansson}(2024)}]{Johansson2024}%
  \BibitemOpen
  \bibfield  {author} {\bibinfo {author} {\bibfnamefont {A.}~\bibnamefont {Johansson}},\ }\bibfield  {title} {\bibinfo {title} {{Theory of spin and orbital Edelstein effects}},\ }\href {https://doi.org/10.1088/1361-648x/ad5e2b} {\bibfield  {journal} {\bibinfo  {journal} {Journal of Physics: Condensed Matter}\ }\textbf {\bibinfo {volume} {36}},\ \bibinfo {pages} {423002} (\bibinfo {year} {2024})}\BibitemShut {NoStop}%
\bibitem [{\citenamefont {Atencia}\ \emph {et~al.}(2024)\citenamefont {Atencia}, \citenamefont {Agarwal},\ and\ \citenamefont {Culcer}}]{BurgosAtencia2024}%
  \BibitemOpen
  \bibfield  {author} {\bibinfo {author} {\bibfnamefont {R.~B.}\ \bibnamefont {Atencia}}, \bibinfo {author} {\bibfnamefont {A.}~\bibnamefont {Agarwal}},\ and\ \bibinfo {author} {\bibfnamefont {D.}~\bibnamefont {Culcer}},\ }\bibfield  {title} {\bibinfo {title} {{Orbital angular momentum of Bloch electrons: equilibrium formulation, magneto-electric phenomena, and the orbital Hall effect}},\ }\href {https://doi.org/10.1080/23746149.2024.2371972} {\bibfield  {journal} {\bibinfo  {journal} {Advances in Physics: X}\ }\textbf {\bibinfo {volume} {9}},\ \bibinfo {pages} {2371972} (\bibinfo {year} {2024})}\BibitemShut {NoStop}%
\bibitem [{\citenamefont {Cysne}\ \emph {et~al.}(2025)\citenamefont {Cysne}, \citenamefont {Canonico}, \citenamefont {Costa}, \citenamefont {Muniz},\ and\ \citenamefont {Rappoport}}]{cysne2025orbitronics}%
  \BibitemOpen
  \bibfield  {author} {\bibinfo {author} {\bibfnamefont {T.~P.}\ \bibnamefont {Cysne}}, \bibinfo {author} {\bibfnamefont {L.~M.}\ \bibnamefont {Canonico}}, \bibinfo {author} {\bibfnamefont {M.}~\bibnamefont {Costa}}, \bibinfo {author} {\bibfnamefont {R.}~\bibnamefont {Muniz}},\ and\ \bibinfo {author} {\bibfnamefont {T.~G.}\ \bibnamefont {Rappoport}},\ }\bibfield  {title} {\bibinfo {title} {{Orbitronics in Two-dimensional Materials}},\ }\bibfield  {journal} {\bibinfo  {journal} {arXiv preprint arXiv:2502.12339}\ }\href {https://doi.org/10.48550/arXiv.2502.12339} {10.48550/arXiv.2502.12339} (\bibinfo {year} {2025})\BibitemShut {NoStop}%
\bibitem [{\citenamefont {Yoda}\ \emph {et~al.}(2015)\citenamefont {Yoda}, \citenamefont {Yokoyama},\ and\ \citenamefont {Murakami}}]{yoda2015current}%
  \BibitemOpen
  \bibfield  {author} {\bibinfo {author} {\bibfnamefont {T.}~\bibnamefont {Yoda}}, \bibinfo {author} {\bibfnamefont {T.}~\bibnamefont {Yokoyama}},\ and\ \bibinfo {author} {\bibfnamefont {S.}~\bibnamefont {Murakami}},\ }\bibfield  {title} {\bibinfo {title} {{Current-induced Orbital and Spin Magnetizations in Crystals with Helical Structure}},\ }\href {https://doi.org/10.1038/srep12024} {\bibfield  {journal} {\bibinfo  {journal} {Scientific Reports}\ }\textbf {\bibinfo {volume} {5}},\ \bibinfo {pages} {12024} (\bibinfo {year} {2015})}\BibitemShut {NoStop}%
\bibitem [{\citenamefont {Yoda}\ \emph {et~al.}(2018)\citenamefont {Yoda}, \citenamefont {Yokoyama},\ and\ \citenamefont {Murakami}}]{yoda2018orbital}%
  \BibitemOpen
  \bibfield  {author} {\bibinfo {author} {\bibfnamefont {T.}~\bibnamefont {Yoda}}, \bibinfo {author} {\bibfnamefont {T.}~\bibnamefont {Yokoyama}},\ and\ \bibinfo {author} {\bibfnamefont {S.}~\bibnamefont {Murakami}},\ }\bibfield  {title} {\bibinfo {title} {{Orbital Edelstein Effect as a Condensed-Matter Analog of Solenoids}},\ }\href {https://doi.org/10.1021/acs.nanolett.7b04300} {\bibfield  {journal} {\bibinfo  {journal} {Nano Letters}\ }\textbf {\bibinfo {volume} {18}},\ \bibinfo {pages} {916} (\bibinfo {year} {2018})}\BibitemShut {NoStop}%
\bibitem [{\citenamefont {Go}\ \emph {et~al.}(2017)\citenamefont {Go}, \citenamefont {Hanke}, \citenamefont {Buhl}, \citenamefont {Freimuth}, \citenamefont {Bihlmayer}, \citenamefont {Lee}, \citenamefont {Mokrousov},\ and\ \citenamefont {Bl\"{u}gel}}]{Go2017}%
  \BibitemOpen
  \bibfield  {author} {\bibinfo {author} {\bibfnamefont {D.}~\bibnamefont {Go}}, \bibinfo {author} {\bibfnamefont {J.-P.}\ \bibnamefont {Hanke}}, \bibinfo {author} {\bibfnamefont {P.~M.}\ \bibnamefont {Buhl}}, \bibinfo {author} {\bibfnamefont {F.}~\bibnamefont {Freimuth}}, \bibinfo {author} {\bibfnamefont {G.}~\bibnamefont {Bihlmayer}}, \bibinfo {author} {\bibfnamefont {H.-W.}\ \bibnamefont {Lee}}, \bibinfo {author} {\bibfnamefont {Y.}~\bibnamefont {Mokrousov}},\ and\ \bibinfo {author} {\bibfnamefont {S.}~\bibnamefont {Bl\"{u}gel}},\ }\bibfield  {title} {\bibinfo {title} {{Toward surface orbitronics: giant orbital magnetism from the orbital Rashba effect at the surface of sp-metals}},\ }\href {https://doi.org/10.1038/srep46742} {\bibfield  {journal} {\bibinfo  {journal} {Scientific Reports}\ }\textbf {\bibinfo {volume} {7}},\ \bibinfo {pages} {46742} (\bibinfo {year} {2017})}\BibitemShut {NoStop}%
\bibitem [{\citenamefont {Salemi}\ \emph {et~al.}(2019)\citenamefont {Salemi}, \citenamefont {Berritta}, \citenamefont {Nandy},\ and\ \citenamefont {Oppeneer}}]{Salemi2019}%
  \BibitemOpen
  \bibfield  {author} {\bibinfo {author} {\bibfnamefont {L.}~\bibnamefont {Salemi}}, \bibinfo {author} {\bibfnamefont {M.}~\bibnamefont {Berritta}}, \bibinfo {author} {\bibfnamefont {A.~K.}\ \bibnamefont {Nandy}},\ and\ \bibinfo {author} {\bibfnamefont {P.~M.}\ \bibnamefont {Oppeneer}},\ }\bibfield  {title} {\bibinfo {title} {{Orbitally dominated Rashba-Edelstein effect in noncentrosymmetric antiferromagnets}},\ }\href {https://doi.org/10.1038/s41467-019-13367-z} {\bibfield  {journal} {\bibinfo  {journal} {Nature Communications}\ }\textbf {\bibinfo {volume} {10}},\ \bibinfo {pages} {5381} (\bibinfo {year} {2019})}\BibitemShut {NoStop}%
\bibitem [{\citenamefont {Nikolaev}\ \emph {et~al.}(2024)\citenamefont {Nikolaev}, \citenamefont {Chshiev}, \citenamefont {Ibrahim}, \citenamefont {Krishnia}, \citenamefont {Sebe}, \citenamefont {George}, \citenamefont {Cros}, \citenamefont {Jaffrès},\ and\ \citenamefont {Fert}}]{Nikolaev2024}%
  \BibitemOpen
  \bibfield  {author} {\bibinfo {author} {\bibfnamefont {S.~A.}\ \bibnamefont {Nikolaev}}, \bibinfo {author} {\bibfnamefont {M.}~\bibnamefont {Chshiev}}, \bibinfo {author} {\bibfnamefont {F.}~\bibnamefont {Ibrahim}}, \bibinfo {author} {\bibfnamefont {S.}~\bibnamefont {Krishnia}}, \bibinfo {author} {\bibfnamefont {N.}~\bibnamefont {Sebe}}, \bibinfo {author} {\bibfnamefont {J.-M.}\ \bibnamefont {George}}, \bibinfo {author} {\bibfnamefont {V.}~\bibnamefont {Cros}}, \bibinfo {author} {\bibfnamefont {H.}~\bibnamefont {Jaffrès}},\ and\ \bibinfo {author} {\bibfnamefont {A.}~\bibnamefont {Fert}},\ }\bibfield  {title} {\bibinfo {title} {{Large Chiral Orbital Texture and Orbital Edelstein Effect in Co/Al Heterostructure}},\ }\href {https://doi.org/10.1021/acs.nanolett.4c01607} {\bibfield  {journal} {\bibinfo  {journal} {Nano Letters}\ }\textbf {\bibinfo {volume} {24}},\ \bibinfo {pages} {13465–} (\bibinfo {year} {2024})}\BibitemShut {NoStop}%
\bibitem [{\citenamefont {Lopez}\ \emph {et~al.}(2012)\citenamefont {Lopez}, \citenamefont {Vanderbilt}, \citenamefont {Thonhauser},\ and\ \citenamefont {Souza}}]{lopez2012wannier}%
  \BibitemOpen
  \bibfield  {author} {\bibinfo {author} {\bibfnamefont {M.~G.}\ \bibnamefont {Lopez}}, \bibinfo {author} {\bibfnamefont {D.}~\bibnamefont {Vanderbilt}}, \bibinfo {author} {\bibfnamefont {T.}~\bibnamefont {Thonhauser}},\ and\ \bibinfo {author} {\bibfnamefont {I.}~\bibnamefont {Souza}},\ }\bibfield  {title} {\bibinfo {title} {{Wannier-based calculation of the orbital magnetization in crystals}},\ }\href {https://doi.org/10.1103/PhysRevB.85.014435} {\bibfield  {journal} {\bibinfo  {journal} {Phys. Rev. B}\ }\textbf {\bibinfo {volume} {85}},\ \bibinfo {pages} {014435} (\bibinfo {year} {2012})}\BibitemShut {NoStop}%
\bibitem [{\citenamefont {Mostofi}\ \emph {et~al.}(2014)\citenamefont {Mostofi}, \citenamefont {Yates}, \citenamefont {Pizzi}, \citenamefont {Lee}, \citenamefont {Souza}, \citenamefont {Vanderbilt},\ and\ \citenamefont {Marzari}}]{mostofi_updated_2014}%
  \BibitemOpen
  \bibfield  {author} {\bibinfo {author} {\bibfnamefont {A.~A.}\ \bibnamefont {Mostofi}}, \bibinfo {author} {\bibfnamefont {J.~R.}\ \bibnamefont {Yates}}, \bibinfo {author} {\bibfnamefont {G.}~\bibnamefont {Pizzi}}, \bibinfo {author} {\bibfnamefont {Y.-S.}\ \bibnamefont {Lee}}, \bibinfo {author} {\bibfnamefont {I.}~\bibnamefont {Souza}}, \bibinfo {author} {\bibfnamefont {D.}~\bibnamefont {Vanderbilt}},\ and\ \bibinfo {author} {\bibfnamefont {N.}~\bibnamefont {Marzari}},\ }\bibfield  {title} {\bibinfo {title} {An updated version of wannier90: {A} tool for obtaining maximally-localised {Wannier} functions},\ }\href {https://doi.org/10.1016/j.cpc.2014.05.003} {\bibfield  {journal} {\bibinfo  {journal} {Computer Physics Communications}\ }\textbf {\bibinfo {volume} {185}},\ \bibinfo {pages} {2309} (\bibinfo {year} {2014})}\BibitemShut {NoStop}%
\bibitem [{\citenamefont {Tsirkin}(2021)}]{Tsirkin2021}%
  \BibitemOpen
  \bibfield  {author} {\bibinfo {author} {\bibfnamefont {S.~S.}\ \bibnamefont {Tsirkin}},\ }\bibfield  {title} {\bibinfo {title} {{High performance Wannier interpolation of Berry curvature and related quantities with WannierBerri code}},\ }\href {https://doi.org/10.1038/s41524-021-00498-5} {\bibfield  {journal} {\bibinfo  {journal} {npj Computational Materials}\ }\textbf {\bibinfo {volume} {7}},\ \bibinfo {pages} {33} (\bibinfo {year} {2021})}\BibitemShut {NoStop}%
\bibitem [{\citenamefont {Krempaský}\ \emph {et~al.}(2016)\citenamefont {Krempaský}, \citenamefont {Muff}, \citenamefont {Bisti}, \citenamefont {Fanciulli}, \citenamefont {Volfová}, \citenamefont {Weber}, \citenamefont {Pilet}, \citenamefont {Warnicke}, \citenamefont {Ebert}, \citenamefont {Braun}, \citenamefont {Bertran}, \citenamefont {Volobuev}, \citenamefont {Minár}, \citenamefont {Springholz}, \citenamefont {Dil},\ and\ \citenamefont {Strocov}}]{Krempasky2016}%
  \BibitemOpen
  \bibfield  {author} {\bibinfo {author} {\bibfnamefont {J.}~\bibnamefont {Krempaský}}, \bibinfo {author} {\bibfnamefont {S.}~\bibnamefont {Muff}}, \bibinfo {author} {\bibfnamefont {F.}~\bibnamefont {Bisti}}, \bibinfo {author} {\bibfnamefont {M.}~\bibnamefont {Fanciulli}}, \bibinfo {author} {\bibfnamefont {H.}~\bibnamefont {Volfová}}, \bibinfo {author} {\bibfnamefont {A.~P.}\ \bibnamefont {Weber}}, \bibinfo {author} {\bibfnamefont {N.}~\bibnamefont {Pilet}}, \bibinfo {author} {\bibfnamefont {P.}~\bibnamefont {Warnicke}}, \bibinfo {author} {\bibfnamefont {H.}~\bibnamefont {Ebert}}, \bibinfo {author} {\bibfnamefont {J.}~\bibnamefont {Braun}}, \bibinfo {author} {\bibfnamefont {F.}~\bibnamefont {Bertran}}, \bibinfo {author} {\bibfnamefont {V.~V.}\ \bibnamefont {Volobuev}}, \bibinfo {author} {\bibfnamefont {J.}~\bibnamefont {Minár}}, \bibinfo {author} {\bibfnamefont {G.}~\bibnamefont {Springholz}}, \bibinfo {author} {\bibfnamefont {J.~H.}\ \bibnamefont {Dil}},\ and\ \bibinfo {author} {\bibfnamefont {V.~N.}\
  \bibnamefont {Strocov}},\ }\bibfield  {title} {\bibinfo {title} {{Entanglement and manipulation of the magnetic and spin–orbit order in multiferroic Rashba semiconductors}},\ }\bibfield  {journal} {\bibinfo  {journal} {Nature Communications}\ }\textbf {\bibinfo {volume} {7}},\ \href {https://doi.org/10.1038/ncomms13071} {10.1038/ncomms13071} (\bibinfo {year} {2016})\BibitemShut {NoStop}%
\bibitem [{\citenamefont {Krempask\'y}\ \emph {et~al.}(2016)\citenamefont {Krempask\'y}, \citenamefont {Volfov\'a}, \citenamefont {Muff}, \citenamefont {Pilet}, \citenamefont {Landolt}, \citenamefont {Radovi\ifmmode~\acute{c}\else \'{c}\fi{}}, \citenamefont {Shi}, \citenamefont {Kriegner}, \citenamefont {Hol\'y}, \citenamefont {Braun}, \citenamefont {Ebert}, \citenamefont {Bisti}, \citenamefont {Rogalev}, \citenamefont {Strocov}, \citenamefont {Springholz}, \citenamefont {Min\'ar},\ and\ \citenamefont {Dil}}]{Krempasky2016_2}%
  \BibitemOpen
  \bibfield  {author} {\bibinfo {author} {\bibfnamefont {J.}~\bibnamefont {Krempask\'y}}, \bibinfo {author} {\bibfnamefont {H.}~\bibnamefont {Volfov\'a}}, \bibinfo {author} {\bibfnamefont {S.}~\bibnamefont {Muff}}, \bibinfo {author} {\bibfnamefont {N.}~\bibnamefont {Pilet}}, \bibinfo {author} {\bibfnamefont {G.}~\bibnamefont {Landolt}}, \bibinfo {author} {\bibfnamefont {M.}~\bibnamefont {Radovi\ifmmode~\acute{c}\else \'{c}\fi{}}}, \bibinfo {author} {\bibfnamefont {M.}~\bibnamefont {Shi}}, \bibinfo {author} {\bibfnamefont {D.}~\bibnamefont {Kriegner}}, \bibinfo {author} {\bibfnamefont {V.}~\bibnamefont {Hol\'y}}, \bibinfo {author} {\bibfnamefont {J.}~\bibnamefont {Braun}}, \bibinfo {author} {\bibfnamefont {H.}~\bibnamefont {Ebert}}, \bibinfo {author} {\bibfnamefont {F.}~\bibnamefont {Bisti}}, \bibinfo {author} {\bibfnamefont {V.~A.}\ \bibnamefont {Rogalev}}, \bibinfo {author} {\bibfnamefont {V.~N.}\ \bibnamefont {Strocov}}, \bibinfo {author} {\bibfnamefont {G.}~\bibnamefont {Springholz}}, \bibinfo {author}
  {\bibfnamefont {J.}~\bibnamefont {Min\'ar}},\ and\ \bibinfo {author} {\bibfnamefont {J.~H.}\ \bibnamefont {Dil}},\ }\bibfield  {title} {\bibinfo {title} {{Disentangling bulk and surface Rashba effects in ferroelectric $\ensuremath{\alpha}$-GeTe}},\ }\href {https://doi.org/10.1103/PhysRevB.94.205111} {\bibfield  {journal} {\bibinfo  {journal} {Phys. Rev. B}\ }\textbf {\bibinfo {volume} {94}},\ \bibinfo {pages} {205111} (\bibinfo {year} {2016})}\BibitemShut {NoStop}%
\bibitem [{\citenamefont {Krempask\'y}\ \emph {et~al.}(2020)\citenamefont {Krempask\'y}, \citenamefont {Fanciulli}, \citenamefont {Nicola\"{\i}}, \citenamefont {Min\'ar}, \citenamefont {Volfov\'a}, \citenamefont {Caha}, \citenamefont {Volobuev}, \citenamefont {S\'anchez-Barriga}, \citenamefont {Gmitra}, \citenamefont {Yaji}, \citenamefont {Kuroda}, \citenamefont {Shin}, \citenamefont {Komori}, \citenamefont {Springholz},\ and\ \citenamefont {Dil}}]{Krempasky2020}%
  \BibitemOpen
  \bibfield  {author} {\bibinfo {author} {\bibfnamefont {J.}~\bibnamefont {Krempask\'y}}, \bibinfo {author} {\bibfnamefont {M.}~\bibnamefont {Fanciulli}}, \bibinfo {author} {\bibfnamefont {L.}~\bibnamefont {Nicola\"{\i}}}, \bibinfo {author} {\bibfnamefont {J.}~\bibnamefont {Min\'ar}}, \bibinfo {author} {\bibfnamefont {H.}~\bibnamefont {Volfov\'a}}, \bibinfo {author} {\bibfnamefont {O.~c.~v.}\ \bibnamefont {Caha}}, \bibinfo {author} {\bibfnamefont {V.~V.}\ \bibnamefont {Volobuev}}, \bibinfo {author} {\bibfnamefont {J.}~\bibnamefont {S\'anchez-Barriga}}, \bibinfo {author} {\bibfnamefont {M.}~\bibnamefont {Gmitra}}, \bibinfo {author} {\bibfnamefont {K.}~\bibnamefont {Yaji}}, \bibinfo {author} {\bibfnamefont {K.}~\bibnamefont {Kuroda}}, \bibinfo {author} {\bibfnamefont {S.}~\bibnamefont {Shin}}, \bibinfo {author} {\bibfnamefont {F.}~\bibnamefont {Komori}}, \bibinfo {author} {\bibfnamefont {G.}~\bibnamefont {Springholz}},\ and\ \bibinfo {author} {\bibfnamefont {J.~H.}\ \bibnamefont {Dil}},\ }\bibfield  {title}
  {\bibinfo {title} {{Fully spin-polarized bulk states in ferroelectric GeTe}},\ }\href {https://doi.org/10.1103/PhysRevResearch.2.013107} {\bibfield  {journal} {\bibinfo  {journal} {Phys. Rev. Res.}\ }\textbf {\bibinfo {volume} {2}},\ \bibinfo {pages} {013107} (\bibinfo {year} {2020})}\BibitemShut {NoStop}%
\bibitem [{\citenamefont {Tenzin}\ \emph {et~al.}(2023)\citenamefont {Tenzin}, \citenamefont {Roy}, \citenamefont {Jafari}, \citenamefont {Banas}, \citenamefont {Cerasoli}, \citenamefont {Date}, \citenamefont {Jayaraj}, \citenamefont {Buongiorno~Nardelli},\ and\ \citenamefont {S\l{}awi\ifmmode~\acute{n}\else \'{n}\fi{}ska}}]{Tenzin2023}%
  \BibitemOpen
  \bibfield  {author} {\bibinfo {author} {\bibfnamefont {K.}~\bibnamefont {Tenzin}}, \bibinfo {author} {\bibfnamefont {A.}~\bibnamefont {Roy}}, \bibinfo {author} {\bibfnamefont {H.}~\bibnamefont {Jafari}}, \bibinfo {author} {\bibfnamefont {B.}~\bibnamefont {Banas}}, \bibinfo {author} {\bibfnamefont {F.~T.}\ \bibnamefont {Cerasoli}}, \bibinfo {author} {\bibfnamefont {M.}~\bibnamefont {Date}}, \bibinfo {author} {\bibfnamefont {A.}~\bibnamefont {Jayaraj}}, \bibinfo {author} {\bibfnamefont {M.}~\bibnamefont {Buongiorno~Nardelli}},\ and\ \bibinfo {author} {\bibfnamefont {J.}~\bibnamefont {S\l{}awi\ifmmode~\acute{n}\else \'{n}\fi{}ska}},\ }\bibfield  {title} {\bibinfo {title} {{Analogs of Rashba-Edelstein effect from density functional theory}},\ }\href {https://doi.org/10.1103/PhysRevB.107.165140} {\bibfield  {journal} {\bibinfo  {journal} {Phys. Rev. B}\ }\textbf {\bibinfo {volume} {107}},\ \bibinfo {pages} {165140} (\bibinfo {year} {2023})}\BibitemShut {NoStop}%
\bibitem [{\citenamefont {Ponet}\ and\ \citenamefont {Artyukhin}(2018)}]{Ponet2018}%
  \BibitemOpen
  \bibfield  {author} {\bibinfo {author} {\bibfnamefont {L.}~\bibnamefont {Ponet}}\ and\ \bibinfo {author} {\bibfnamefont {S.}~\bibnamefont {Artyukhin}},\ }\bibfield  {title} {\bibinfo {title} {{First-principles theory of giant Rashba-like spin splitting in bulk GeTe}},\ }\href {https://doi.org/10.1103/PhysRevB.98.174102} {\bibfield  {journal} {\bibinfo  {journal} {Phys. Rev. B}\ }\textbf {\bibinfo {volume} {98}},\ \bibinfo {pages} {174102} (\bibinfo {year} {2018})}\BibitemShut {NoStop}%
\bibitem [{\citenamefont {Kresse}\ and\ \citenamefont {Hafner}(1993)}]{Kresse_1993}%
  \BibitemOpen
  \bibfield  {author} {\bibinfo {author} {\bibfnamefont {G.}~\bibnamefont {Kresse}}\ and\ \bibinfo {author} {\bibfnamefont {J.}~\bibnamefont {Hafner}},\ }\bibfield  {title} {\bibinfo {title} {{Ab initio molecular dynamics for liquid metals}},\ }\href {https://doi.org/10.1103/PhysRevB.47.558} {\bibfield  {journal} {\bibinfo  {journal} {Phys. Rev. B}\ }\textbf {\bibinfo {volume} {47}},\ \bibinfo {pages} {558} (\bibinfo {year} {1993})}\BibitemShut {NoStop}%
\bibitem [{\citenamefont {Kresse}\ and\ \citenamefont {Furthm\"uller}(1996)}]{Kresse_1996a}%
  \BibitemOpen
  \bibfield  {author} {\bibinfo {author} {\bibfnamefont {G.}~\bibnamefont {Kresse}}\ and\ \bibinfo {author} {\bibfnamefont {J.}~\bibnamefont {Furthm\"uller}},\ }\bibfield  {title} {\bibinfo {title} {{Efficient iterative schemes for ab initio total-energy calculations using a plane-wave basis set}},\ }\href {https://doi.org/10.1103/PhysRevB.54.11169} {\bibfield  {journal} {\bibinfo  {journal} {Phys. Rev. B}\ }\textbf {\bibinfo {volume} {54}},\ \bibinfo {pages} {11169} (\bibinfo {year} {1996})}\BibitemShut {NoStop}%
\bibitem [{\citenamefont {Kresse}\ and\ \citenamefont {Furthmüller}(1996)}]{Kresse_1996b}%
  \BibitemOpen
  \bibfield  {author} {\bibinfo {author} {\bibfnamefont {G.}~\bibnamefont {Kresse}}\ and\ \bibinfo {author} {\bibfnamefont {J.}~\bibnamefont {Furthmüller}},\ }\bibfield  {title} {\bibinfo {title} {{Efficiency of ab-initio total energy calculations for metals and semiconductors using a plane-wave basis set}},\ }\href {https://doi.org/https://doi.org/10.1016/0927-0256(96)00008-0} {\bibfield  {journal} {\bibinfo  {journal} {Computational Materials Science}\ }\textbf {\bibinfo {volume} {6}},\ \bibinfo {pages} {15} (\bibinfo {year} {1996})}\BibitemShut {NoStop}%
\bibitem [{\citenamefont {Kresse}\ and\ \citenamefont {Joubert}(1999)}]{Kresse_1999}%
  \BibitemOpen
  \bibfield  {author} {\bibinfo {author} {\bibfnamefont {G.}~\bibnamefont {Kresse}}\ and\ \bibinfo {author} {\bibfnamefont {D.}~\bibnamefont {Joubert}},\ }\bibfield  {title} {\bibinfo {title} {{From ultrasoft pseudopotentials to the projector augmented-wave method}},\ }\href {https://doi.org/10.1103/PhysRevB.59.1758} {\bibfield  {journal} {\bibinfo  {journal} {Phys. Rev. B}\ }\textbf {\bibinfo {volume} {59}},\ \bibinfo {pages} {1758} (\bibinfo {year} {1999})}\BibitemShut {NoStop}%
\bibitem [{\citenamefont {Perdew}\ \emph {et~al.}(1996)\citenamefont {Perdew}, \citenamefont {Burke},\ and\ \citenamefont {Ernzerhof}}]{Perdew_1996}%
  \BibitemOpen
  \bibfield  {author} {\bibinfo {author} {\bibfnamefont {J.~P.}\ \bibnamefont {Perdew}}, \bibinfo {author} {\bibfnamefont {K.}~\bibnamefont {Burke}},\ and\ \bibinfo {author} {\bibfnamefont {M.}~\bibnamefont {Ernzerhof}},\ }\bibfield  {title} {\bibinfo {title} {{Generalized Gradient Approximation Made Simple}},\ }\href {https://doi.org/10.1103/PhysRevLett.77.3865} {\bibfield  {journal} {\bibinfo  {journal} {Phys. Rev. Lett.}\ }\textbf {\bibinfo {volume} {77}},\ \bibinfo {pages} {3865} (\bibinfo {year} {1996})}\BibitemShut {NoStop}%
\bibitem [{\citenamefont {Grimme}\ \emph {et~al.}(2010)\citenamefont {Grimme}, \citenamefont {Antony}, \citenamefont {Ehrlich},\ and\ \citenamefont {Krieg}}]{Grimme_2010}%
  \BibitemOpen
  \bibfield  {author} {\bibinfo {author} {\bibfnamefont {S.}~\bibnamefont {Grimme}}, \bibinfo {author} {\bibfnamefont {J.}~\bibnamefont {Antony}}, \bibinfo {author} {\bibfnamefont {S.}~\bibnamefont {Ehrlich}},\ and\ \bibinfo {author} {\bibfnamefont {H.}~\bibnamefont {Krieg}},\ }\bibfield  {title} {\bibinfo {title} {{A consistent and accurate ab initio parametrization of density functional dispersion correction (DFT-D) for the 94 elements H-Pu}},\ }\href {https://doi.org/10.1063/1.3382344} {\bibfield  {journal} {\bibinfo  {journal} {The Journal of Chemical Physics}\ }\textbf {\bibinfo {volume} {132}},\ \bibinfo {pages} {154104} (\bibinfo {year} {2010})}\BibitemShut {NoStop}%
\bibitem [{\citenamefont {Grimme}\ \emph {et~al.}(2011)\citenamefont {Grimme}, \citenamefont {Ehrlich},\ and\ \citenamefont {Goerigk}}]{Grimme2011}%
  \BibitemOpen
  \bibfield  {author} {\bibinfo {author} {\bibfnamefont {S.}~\bibnamefont {Grimme}}, \bibinfo {author} {\bibfnamefont {S.}~\bibnamefont {Ehrlich}},\ and\ \bibinfo {author} {\bibfnamefont {L.}~\bibnamefont {Goerigk}},\ }\bibfield  {title} {\bibinfo {title} {{Effect of the damping function in dispersion corrected density functional theory}},\ }\href {https://doi.org/10.1002/jcc.21759} {\bibfield  {journal} {\bibinfo  {journal} {Journal of Computational Chemistry}\ }\textbf {\bibinfo {volume} {32}},\ \bibinfo {pages} {1456–1465} (\bibinfo {year} {2011})}\BibitemShut {NoStop}%
\bibitem [{\citenamefont {Marzari}\ and\ \citenamefont {Vanderbilt}(1997)}]{marzari_maximally_1997}%
  \BibitemOpen
  \bibfield  {author} {\bibinfo {author} {\bibfnamefont {N.}~\bibnamefont {Marzari}}\ and\ \bibinfo {author} {\bibfnamefont {D.}~\bibnamefont {Vanderbilt}},\ }\bibfield  {title} {\bibinfo {title} {Maximally localized generalized {Wannier} functions for composite energy bands},\ }\href {https://doi.org/10.1103/PhysRevB.56.12847} {\bibfield  {journal} {\bibinfo  {journal} {Physical Review B}\ }\textbf {\bibinfo {volume} {56}},\ \bibinfo {pages} {12847} (\bibinfo {year} {1997})}\BibitemShut {NoStop}%
\bibitem [{\citenamefont {Marzari}\ \emph {et~al.}(2012)\citenamefont {Marzari}, \citenamefont {Mostofi}, \citenamefont {Yates}, \citenamefont {Souza},\ and\ \citenamefont {Vanderbilt}}]{marzari_maximally_2012}%
  \BibitemOpen
  \bibfield  {author} {\bibinfo {author} {\bibfnamefont {N.}~\bibnamefont {Marzari}}, \bibinfo {author} {\bibfnamefont {A.~A.}\ \bibnamefont {Mostofi}}, \bibinfo {author} {\bibfnamefont {J.~R.}\ \bibnamefont {Yates}}, \bibinfo {author} {\bibfnamefont {I.}~\bibnamefont {Souza}},\ and\ \bibinfo {author} {\bibfnamefont {D.}~\bibnamefont {Vanderbilt}},\ }\bibfield  {title} {\bibinfo {title} {Maximally localized {Wannier} functions: {Theory} and applications},\ }\href {https://doi.org/10.1103/RevModPhys.84.1419} {\bibfield  {journal} {\bibinfo  {journal} {Reviews of Modern Physics}\ }\textbf {\bibinfo {volume} {84}},\ \bibinfo {pages} {1419} (\bibinfo {year} {2012})}\BibitemShut {NoStop}%
\bibitem [{\citenamefont {Vojáček}\ \emph {et~al.}(2024{\natexlab{b}})\citenamefont {Vojáček}, \citenamefont {Medina~Dueñas}, \citenamefont {Li}, \citenamefont {Ibrahim}, \citenamefont {Manchon}, \citenamefont {Roche}, \citenamefont {Chshiev},\ and\ \citenamefont {García}}]{vojacek_field-free_2024}%
  \BibitemOpen
  \bibfield  {author} {\bibinfo {author} {\bibfnamefont {L.}~\bibnamefont {Vojáček}}, \bibinfo {author} {\bibfnamefont {J.}~\bibnamefont {Medina~Dueñas}}, \bibinfo {author} {\bibfnamefont {J.}~\bibnamefont {Li}}, \bibinfo {author} {\bibfnamefont {F.}~\bibnamefont {Ibrahim}}, \bibinfo {author} {\bibfnamefont {A.}~\bibnamefont {Manchon}}, \bibinfo {author} {\bibfnamefont {S.}~\bibnamefont {Roche}}, \bibinfo {author} {\bibfnamefont {M.}~\bibnamefont {Chshiev}},\ and\ \bibinfo {author} {\bibfnamefont {J.~H.}\ \bibnamefont {García}},\ }\bibfield  {title} {\bibinfo {title} {Field-{Free} {Spin}–{Orbit} {Torque} {Switching} in {Janus} {Chromium} {Dichalcogenides}},\ }\href {https://doi.org/10.1021/acs.nanolett.4c03029} {\bibfield  {journal} {\bibinfo  {journal} {Nano Letters}\ }\textbf {\bibinfo {volume} {24}},\ \bibinfo {pages} {11889} (\bibinfo {year} {2024}{\natexlab{b}})}\BibitemShut {NoStop}%
\bibitem [{\citenamefont {Vojáček}(2024)}]{vojacek_multiscale_2024}%
  \BibitemOpen
  \bibfield  {author} {\bibinfo {author} {\bibfnamefont {L.}~\bibnamefont {Vojáček}},\ }\emph {\bibinfo {title} {Multiscale modeling of spin-orbitronic phenomena at metal, oxide, and {2D} material interfaces for spintronic devices}},\ \href {https://theses.hal.science/tel-04836940} {Ph.D. thesis},\ \bibinfo  {school} {Université Grenoble Alpes} (\bibinfo {year} {2024})\BibitemShut {NoStop}%
\bibitem [{\citenamefont {Fu}(2009)}]{Fu2009}%
  \BibitemOpen
  \bibfield  {author} {\bibinfo {author} {\bibfnamefont {L.}~\bibnamefont {Fu}},\ }\bibfield  {title} {\bibinfo {title} {{Hexagonal Warping Effects in the Surface States of the Topological Insulator ${\mathrm{Bi}}_{2}{\mathrm{Te}}_{3}$}},\ }\href {https://doi.org/10.1103/PhysRevLett.103.266801} {\bibfield  {journal} {\bibinfo  {journal} {Phys. Rev. Lett.}\ }\textbf {\bibinfo {volume} {103}},\ \bibinfo {pages} {266801} (\bibinfo {year} {2009})}\BibitemShut {NoStop}%
\bibitem [{\citenamefont {Cysne}\ \emph {et~al.}(2024)\citenamefont {Cysne}, \citenamefont {Kort-Kamp},\ and\ \citenamefont {Rappoport}}]{cysne2024controlling}%
  \BibitemOpen
  \bibfield  {author} {\bibinfo {author} {\bibfnamefont {T.~P.}\ \bibnamefont {Cysne}}, \bibinfo {author} {\bibfnamefont {W.~J.~M.}\ \bibnamefont {Kort-Kamp}},\ and\ \bibinfo {author} {\bibfnamefont {T.~G.}\ \bibnamefont {Rappoport}},\ }\bibfield  {title} {\bibinfo {title} {{Controlling the orbital Hall effect in gapped bilayer graphene in the terahertz regime}},\ }\href {https://doi.org/10.1103/PhysRevResearch.6.023271} {\bibfield  {journal} {\bibinfo  {journal} {Phys. Rev. Res.}\ }\textbf {\bibinfo {volume} {6}},\ \bibinfo {pages} {023271} (\bibinfo {year} {2024})}\BibitemShut {NoStop}%
\bibitem [{\citenamefont {Hakioglu}\ \emph {et~al.}(2023)\citenamefont {Hakioglu}, \citenamefont {Chiu}, \citenamefont {Markiewicz}, \citenamefont {Singh},\ and\ \citenamefont {Bansil}}]{Hakioglu2023}%
  \BibitemOpen
  \bibfield  {author} {\bibinfo {author} {\bibfnamefont {T.}~\bibnamefont {Hakioglu}}, \bibinfo {author} {\bibfnamefont {W.-C.}\ \bibnamefont {Chiu}}, \bibinfo {author} {\bibfnamefont {R.~S.}\ \bibnamefont {Markiewicz}}, \bibinfo {author} {\bibfnamefont {B.}~\bibnamefont {Singh}},\ and\ \bibinfo {author} {\bibfnamefont {A.}~\bibnamefont {Bansil}},\ }\bibfield  {title} {\bibinfo {title} {{Nonorthogonal spin-momentum locking}},\ }\href {https://doi.org/10.1103/PhysRevB.108.155103} {\bibfield  {journal} {\bibinfo  {journal} {Phys. Rev. B}\ }\textbf {\bibinfo {volume} {108}},\ \bibinfo {pages} {155103} (\bibinfo {year} {2023})}\BibitemShut {NoStop}%
\bibitem [{\citenamefont {Kadlec}\ \emph {et~al.}(2011)\citenamefont {Kadlec}, \citenamefont {Kadlec}, \citenamefont {Ku\ifmmode~\check{z}\else \v{z}\fi{}el},\ and\ \citenamefont {Petzelt}}]{Kadlec2011}%
  \BibitemOpen
  \bibfield  {author} {\bibinfo {author} {\bibfnamefont {F.}~\bibnamefont {Kadlec}}, \bibinfo {author} {\bibfnamefont {C.}~\bibnamefont {Kadlec}}, \bibinfo {author} {\bibfnamefont {P.}~\bibnamefont {Ku\ifmmode~\check{z}\else \v{z}\fi{}el}},\ and\ \bibinfo {author} {\bibfnamefont {J.}~\bibnamefont {Petzelt}},\ }\bibfield  {title} {\bibinfo {title} {{Study of the ferroelectric phase transition in germanium telluride using time-domain terahertz spectroscopy}},\ }\href {https://doi.org/10.1103/PhysRevB.84.205209} {\bibfield  {journal} {\bibinfo  {journal} {Phys. Rev. B}\ }\textbf {\bibinfo {volume} {84}},\ \bibinfo {pages} {205209} (\bibinfo {year} {2011})}\BibitemShut {NoStop}%
\bibitem [{\citenamefont {Boschker}\ \emph {et~al.}(2018)\citenamefont {Boschker}, \citenamefont {Lü}, \citenamefont {Bragaglia}, \citenamefont {Wang}, \citenamefont {Grahn},\ and\ \citenamefont {Calarco}}]{Boschker2018}%
  \BibitemOpen
  \bibfield  {author} {\bibinfo {author} {\bibfnamefont {J.~E.}\ \bibnamefont {Boschker}}, \bibinfo {author} {\bibfnamefont {X.}~\bibnamefont {Lü}}, \bibinfo {author} {\bibfnamefont {V.}~\bibnamefont {Bragaglia}}, \bibinfo {author} {\bibfnamefont {R.}~\bibnamefont {Wang}}, \bibinfo {author} {\bibfnamefont {H.~T.}\ \bibnamefont {Grahn}},\ and\ \bibinfo {author} {\bibfnamefont {R.}~\bibnamefont {Calarco}},\ }\bibfield  {title} {\bibinfo {title} {{Electrical and optical properties of epitaxial binary and ternary GeTe-Sb$_2$Te$_3$ alloys}},\ }\href {https://doi.org/10.1038/s41598-018-23221-9} {\bibfield  {journal} {\bibinfo  {journal} {Scientific Reports}\ }\textbf {\bibinfo {volume} {8}},\ \bibinfo {pages} {5889} (\bibinfo {year} {2018})}\BibitemShut {NoStop}%
\bibitem [{\citenamefont {Guo}\ \emph {et~al.}(2019)\citenamefont {Guo}, \citenamefont {Li}, \citenamefont {Qiu}, \citenamefont {Yang}, \citenamefont {Li}, \citenamefont {Shao}, \citenamefont {Chen}, \citenamefont {Ma}, \citenamefont {Sun}, \citenamefont {Cao}, \citenamefont {Zeng}, \citenamefont {Wang},\ and\ \citenamefont {Xie}}]{Guo2019}%
  \BibitemOpen
  \bibfield  {author} {\bibinfo {author} {\bibfnamefont {D.}~\bibnamefont {Guo}}, \bibinfo {author} {\bibfnamefont {C.}~\bibnamefont {Li}}, \bibinfo {author} {\bibfnamefont {K.}~\bibnamefont {Qiu}}, \bibinfo {author} {\bibfnamefont {Q.}~\bibnamefont {Yang}}, \bibinfo {author} {\bibfnamefont {K.}~\bibnamefont {Li}}, \bibinfo {author} {\bibfnamefont {B.}~\bibnamefont {Shao}}, \bibinfo {author} {\bibfnamefont {D.}~\bibnamefont {Chen}}, \bibinfo {author} {\bibfnamefont {Y.}~\bibnamefont {Ma}}, \bibinfo {author} {\bibfnamefont {J.}~\bibnamefont {Sun}}, \bibinfo {author} {\bibfnamefont {X.}~\bibnamefont {Cao}}, \bibinfo {author} {\bibfnamefont {W.}~\bibnamefont {Zeng}}, \bibinfo {author} {\bibfnamefont {Z.}~\bibnamefont {Wang}},\ and\ \bibinfo {author} {\bibfnamefont {R.}~\bibnamefont {Xie}},\ }\bibfield  {title} {\bibinfo {title} {{The n- and p-type thermoelectricity property of GeTe by first-principles study}},\ }\href {https://doi.org/https://doi.org/10.1016/j.jallcom.2019.151838} {\bibfield  {journal} {\bibinfo
   {journal} {Journal of Alloys and Compounds}\ }\textbf {\bibinfo {volume} {810}},\ \bibinfo {pages} {151838} (\bibinfo {year} {2019})}\BibitemShut {NoStop}%
\bibitem [{\citenamefont {Vojáček}\ \emph {et~al.}(2024{\natexlab{c}})\citenamefont {Vojáček}, \citenamefont {Chshiev},\ and\ \citenamefont {Li}}]{vojacek_domain_2024}%
  \BibitemOpen
  \bibfield  {author} {\bibinfo {author} {\bibfnamefont {L.}~\bibnamefont {Vojáček}}, \bibinfo {author} {\bibfnamefont {M.}~\bibnamefont {Chshiev}},\ and\ \bibinfo {author} {\bibfnamefont {J.}~\bibnamefont {Li}},\ }\bibfield  {title} {\bibinfo {title} {Domain {Wall} {Migration}-{Mediated} {Ferroelectric} {Switching} and {Rashba} {Effect} {Tuning} in {GeTe} {Thin} {Films}},\ }\href {https://doi.org/10.1021/acsaelm.4c00392} {\bibfield  {journal} {\bibinfo  {journal} {ACS Applied Electronic Materials}\ }\textbf {\bibinfo {volume} {6}},\ \bibinfo {pages} {3754} (\bibinfo {year} {2024}{\natexlab{c}})},\ \bibinfo {note} {publisher: American Chemical Society}\BibitemShut {NoStop}%
\bibitem [{\citenamefont {Krempask\'y}\ \emph {et~al.}(2018)\citenamefont {Krempask\'y}, \citenamefont {Muff}, \citenamefont {Min\'ar}, \citenamefont {Pilet}, \citenamefont {Fanciulli}, \citenamefont {Weber}, \citenamefont {Guedes}, \citenamefont {Caputo}, \citenamefont {M\"uller}, \citenamefont {Volobuev}, \citenamefont {Gmitra}, \citenamefont {Vaz}, \citenamefont {Scagnoli}, \citenamefont {Springholz},\ and\ \citenamefont {Dil}}]{Krempasky2018}%
  \BibitemOpen
  \bibfield  {author} {\bibinfo {author} {\bibfnamefont {J.}~\bibnamefont {Krempask\'y}}, \bibinfo {author} {\bibfnamefont {S.}~\bibnamefont {Muff}}, \bibinfo {author} {\bibfnamefont {J.}~\bibnamefont {Min\'ar}}, \bibinfo {author} {\bibfnamefont {N.}~\bibnamefont {Pilet}}, \bibinfo {author} {\bibfnamefont {M.}~\bibnamefont {Fanciulli}}, \bibinfo {author} {\bibfnamefont {A.~P.}\ \bibnamefont {Weber}}, \bibinfo {author} {\bibfnamefont {E.~B.}\ \bibnamefont {Guedes}}, \bibinfo {author} {\bibfnamefont {M.}~\bibnamefont {Caputo}}, \bibinfo {author} {\bibfnamefont {E.}~\bibnamefont {M\"uller}}, \bibinfo {author} {\bibfnamefont {V.~V.}\ \bibnamefont {Volobuev}}, \bibinfo {author} {\bibfnamefont {M.}~\bibnamefont {Gmitra}}, \bibinfo {author} {\bibfnamefont {C.~A.~F.}\ \bibnamefont {Vaz}}, \bibinfo {author} {\bibfnamefont {V.}~\bibnamefont {Scagnoli}}, \bibinfo {author} {\bibfnamefont {G.}~\bibnamefont {Springholz}},\ and\ \bibinfo {author} {\bibfnamefont {J.~H.}\ \bibnamefont {Dil}},\ }\bibfield  {title} {\bibinfo
  {title} {{Operando Imaging of All-Electric Spin Texture Manipulation in Ferroelectric and Multiferroic Rashba Semiconductors}},\ }\href {https://doi.org/10.1103/PhysRevX.8.021067} {\bibfield  {journal} {\bibinfo  {journal} {Phys. Rev. X}\ }\textbf {\bibinfo {volume} {8}},\ \bibinfo {pages} {021067} (\bibinfo {year} {2018})}\BibitemShut {NoStop}%
\bibitem [{\citenamefont {Yang}\ \emph {et~al.}(2021{\natexlab{b}})\citenamefont {Yang}, \citenamefont {Li}, \citenamefont {Li}, \citenamefont {Li}, \citenamefont {Sun}, \citenamefont {Liu}, \citenamefont {Bai}, \citenamefont {Li}, \citenamefont {Xie}, \citenamefont {Su}, \citenamefont {Gong}, \citenamefont {Zhang}, \citenamefont {He},\ and\ \citenamefont {Cheng}}]{Yang_2021}%
  \BibitemOpen
  \bibfield  {author} {\bibinfo {author} {\bibfnamefont {X.}~\bibnamefont {Yang}}, \bibinfo {author} {\bibfnamefont {X.-M.}\ \bibnamefont {Li}}, \bibinfo {author} {\bibfnamefont {Y.}~\bibnamefont {Li}}, \bibinfo {author} {\bibfnamefont {Y.}~\bibnamefont {Li}}, \bibinfo {author} {\bibfnamefont {R.}~\bibnamefont {Sun}}, \bibinfo {author} {\bibfnamefont {J.-N.}\ \bibnamefont {Liu}}, \bibinfo {author} {\bibfnamefont {X.}~\bibnamefont {Bai}}, \bibinfo {author} {\bibfnamefont {N.}~\bibnamefont {Li}}, \bibinfo {author} {\bibfnamefont {Z.-K.}\ \bibnamefont {Xie}}, \bibinfo {author} {\bibfnamefont {L.}~\bibnamefont {Su}}, \bibinfo {author} {\bibfnamefont {Z.-Z.}\ \bibnamefont {Gong}}, \bibinfo {author} {\bibfnamefont {X.-Q.}\ \bibnamefont {Zhang}}, \bibinfo {author} {\bibfnamefont {W.}~\bibnamefont {He}},\ and\ \bibinfo {author} {\bibfnamefont {Z.}~\bibnamefont {Cheng}},\ }\bibfield  {title} {\bibinfo {title} {{Three-Dimensional Limit of Bulk Rashba Effect in Ferroelectric Semiconductor GeTe}},\ }\href
  {https://doi.org/10.1021/acs.nanolett.0c03161} {\bibfield  {journal} {\bibinfo  {journal} {Nano Letters}\ }\textbf {\bibinfo {volume} {21}},\ \bibinfo {pages} {77} (\bibinfo {year} {2021}{\natexlab{b}})}\BibitemShut {NoStop}%
\bibitem [{\citenamefont {Deringer}\ \emph {et~al.}(2012)\citenamefont {Deringer}, \citenamefont {Lumeij},\ and\ \citenamefont {Dronskowski}}]{Deringer_2012}%
  \BibitemOpen
  \bibfield  {author} {\bibinfo {author} {\bibfnamefont {V.~L.}\ \bibnamefont {Deringer}}, \bibinfo {author} {\bibfnamefont {M.}~\bibnamefont {Lumeij}},\ and\ \bibinfo {author} {\bibfnamefont {R.}~\bibnamefont {Dronskowski}},\ }\bibfield  {title} {\bibinfo {title} {{Ab Initio Modeling of $\alpha$-GeTe(111) Surfaces}},\ }\href {https://doi.org/10.1021/jp304455z} {\bibfield  {journal} {\bibinfo  {journal} {The Journal of Physical Chemistry C}\ }\textbf {\bibinfo {volume} {116}},\ \bibinfo {pages} {15801} (\bibinfo {year} {2012})}\BibitemShut {NoStop}%
\bibitem [{\citenamefont {Jafari}\ \emph {et~al.}(2022)\citenamefont {Jafari}, \citenamefont {Roy},\ and\ \citenamefont {S\l{}awi\ifmmode~\acute{n}\else \'{n}\fi{}ska}}]{jafari2022ferroelectric}%
  \BibitemOpen
  \bibfield  {author} {\bibinfo {author} {\bibfnamefont {H.}~\bibnamefont {Jafari}}, \bibinfo {author} {\bibfnamefont {A.}~\bibnamefont {Roy}},\ and\ \bibinfo {author} {\bibfnamefont {J.}~\bibnamefont {S\l{}awi\ifmmode~\acute{n}\else \'{n}\fi{}ska}},\ }\bibfield  {title} {\bibinfo {title} {{Ferroelectric control of charge-to-spin conversion in ${\mathrm{WTe}}_{2}$}},\ }\href {https://doi.org/10.1103/PhysRevMaterials.6.L091404} {\bibfield  {journal} {\bibinfo  {journal} {Phys. Rev. Mater.}\ }\textbf {\bibinfo {volume} {6}},\ \bibinfo {pages} {L091404} (\bibinfo {year} {2022})}\BibitemShut {NoStop}%
\bibitem [{\citenamefont {Souza}\ \emph {et~al.}(2024)\citenamefont {Souza}, \citenamefont {Abr\~ao}, \citenamefont {Correa}, \citenamefont {Bohn}, \citenamefont {Gamino},\ and\ \citenamefont {Rezende}}]{souza2024unveiling}%
  \BibitemOpen
  \bibfield  {author} {\bibinfo {author} {\bibfnamefont {E.~C.}\ \bibnamefont {Souza}}, \bibinfo {author} {\bibfnamefont {J.~E.}\ \bibnamefont {Abr\~ao}}, \bibinfo {author} {\bibfnamefont {M.~A.}\ \bibnamefont {Correa}}, \bibinfo {author} {\bibfnamefont {F.}~\bibnamefont {Bohn}}, \bibinfo {author} {\bibfnamefont {M.}~\bibnamefont {Gamino}},\ and\ \bibinfo {author} {\bibfnamefont {S.~M.}\ \bibnamefont {Rezende}},\ }\bibfield  {title} {\bibinfo {title} {Unveiling the mechanism of spin to charge conversion in the ferroelectric topological crystalline insulator snte},\ }\href {https://doi.org/10.1103/PhysRevB.110.014444} {\bibfield  {journal} {\bibinfo  {journal} {Phys. Rev. B}\ }\textbf {\bibinfo {volume} {110}},\ \bibinfo {pages} {014444} (\bibinfo {year} {2024})}\BibitemShut {NoStop}%
\bibitem [{\citenamefont {Go}\ \emph {et~al.}(2018)\citenamefont {Go}, \citenamefont {Jo}, \citenamefont {Kim},\ and\ \citenamefont {Lee}}]{Go2018}%
  \BibitemOpen
  \bibfield  {author} {\bibinfo {author} {\bibfnamefont {D.}~\bibnamefont {Go}}, \bibinfo {author} {\bibfnamefont {D.}~\bibnamefont {Jo}}, \bibinfo {author} {\bibfnamefont {C.}~\bibnamefont {Kim}},\ and\ \bibinfo {author} {\bibfnamefont {H.-W.}\ \bibnamefont {Lee}},\ }\bibfield  {title} {\bibinfo {title} {{Intrinsic Spin and Orbital Hall Effects from Orbital Texture}},\ }\href {https://doi.org/10.1103/PhysRevLett.121.086602} {\bibfield  {journal} {\bibinfo  {journal} {Phys. Rev. Lett.}\ }\textbf {\bibinfo {volume} {121}},\ \bibinfo {pages} {086602} (\bibinfo {year} {2018})}\BibitemShut {NoStop}%
\bibitem [{\citenamefont {Leiva}\ \emph {et~al.}(2025)\citenamefont {Leiva}, \citenamefont {Vojacek}, \citenamefont {Li}, \citenamefont {Chshiev}, \citenamefont {Vila}, \citenamefont {Mertig},\ and\ \citenamefont {Johansson}}]{leiva2025data}%
  \BibitemOpen
  \bibfield  {author} {\bibinfo {author} {\bibfnamefont {S.}~\bibnamefont {Leiva}}, \bibinfo {author} {\bibfnamefont {L.}~\bibnamefont {Vojacek}}, \bibinfo {author} {\bibfnamefont {J.}~\bibnamefont {Li}}, \bibinfo {author} {\bibfnamefont {M.}~\bibnamefont {Chshiev}}, \bibinfo {author} {\bibfnamefont {L.}~\bibnamefont {Vila}}, \bibinfo {author} {\bibfnamefont {I.}~\bibnamefont {Mertig}},\ and\ \bibinfo {author} {\bibfnamefont {A.}~\bibnamefont {Johansson}},\ }\href {https://doi.org/10.5281/zenodo.15536123} {\bibinfo {title} {{Supporting data for "Current-induced spin and orbital polarization in the ferroelectric Rashba semiconductor GeTe"}}} (\bibinfo {year} {2025}),\ \bibinfo {note} {{[Data set], Zenodo, https://doi.org/10.5281/zenodo.15536123}}\BibitemShut {NoStop}%
\bibitem [{\citenamefont {Kremer}\ \emph {et~al.}(2020)\citenamefont {Kremer}, \citenamefont {Jaouen}, \citenamefont {Salzmann}, \citenamefont {Nicola\"{\i}}, \citenamefont {Rumo}, \citenamefont {Nicholson}, \citenamefont {Hildebrand}, \citenamefont {Dil}, \citenamefont {Min\'ar}, \citenamefont {Springholz}, \citenamefont {Krempask\'y},\ and\ \citenamefont {Monney}}]{Kremer2020}%
  \BibitemOpen
  \bibfield  {author} {\bibinfo {author} {\bibfnamefont {G.}~\bibnamefont {Kremer}}, \bibinfo {author} {\bibfnamefont {T.}~\bibnamefont {Jaouen}}, \bibinfo {author} {\bibfnamefont {B.}~\bibnamefont {Salzmann}}, \bibinfo {author} {\bibfnamefont {L.}~\bibnamefont {Nicola\"{\i}}}, \bibinfo {author} {\bibfnamefont {M.}~\bibnamefont {Rumo}}, \bibinfo {author} {\bibfnamefont {C.~W.}\ \bibnamefont {Nicholson}}, \bibinfo {author} {\bibfnamefont {B.}~\bibnamefont {Hildebrand}}, \bibinfo {author} {\bibfnamefont {J.~H.}\ \bibnamefont {Dil}}, \bibinfo {author} {\bibfnamefont {J.}~\bibnamefont {Min\'ar}}, \bibinfo {author} {\bibfnamefont {G.}~\bibnamefont {Springholz}}, \bibinfo {author} {\bibfnamefont {J.}~\bibnamefont {Krempask\'y}},\ and\ \bibinfo {author} {\bibfnamefont {C.}~\bibnamefont {Monney}},\ }\bibfield  {title} {\bibinfo {title} {{Unveiling the complete dispersion of the giant Rashba split surface states of ferroelectric $\ensuremath{\alpha}\text{\ensuremath{-}}\mathrm{GeTe}(111)$ by alkali doping}},\ }\href
  {https://doi.org/10.1103/PhysRevResearch.2.033115} {\bibfield  {journal} {\bibinfo  {journal} {Phys. Rev. Res.}\ }\textbf {\bibinfo {volume} {2}},\ \bibinfo {pages} {033115} (\bibinfo {year} {2020})}\BibitemShut {NoStop}%
\end{thebibliography}%
\end{document}